\documentclass[twocolumn,showpacs,amsmath,amssymb]{revtex4}
\usepackage{graphicx}
\usepackage{dcolumn}
\usepackage{bm}
\newcommand{\beq}{\begin{equation}}
\newcommand{\eeq}{\end{equation}}
\newcommand{\bea}{\begin{eqnarray}}
\newcommand{\eea}{\end{eqnarray}}
\newcommand{\eps}{\varepsilon}
\newcommand{\bfg}{\boldsymbol}

\begin{document}

\author{N. V. Gnezdilov}
\affiliation{Kurchatov Institute, 123182 Moscow, Russia}
\affiliation{National Research Nuclear University MEPhI, 115409
Moscow, Russia}

\author{I. N. Borzov}
\affiliation{Institute for Physics and Power Engineering, 249033
Obninsk, Russia} \affiliation{Joint Institute for Nuclear Research,
141980 Dubna, Russia}

\author{E. E. Saperstein}
\affiliation{Kurchatov Institute, 123182 Moscow, Russia}

\author{S. V. Tolokonnikov}
\affiliation{Kurchatov Institute, 123182 Moscow, Russia}
\affiliation{Moscow Institute of Physics and Technology, 141700
Dolgoprudny, Russia.}

\title{Self-consistent description of single-particle levels of magic nuclei.}


\begin{abstract}
Single-particle levels of seven magic nuclei are calculated within
the Energy Density Functional (EDF) method by Fayans {\it et al.} 
[Nucl. Phys. A {\bf 676}, 49 (2000)]. 
Three versions of the EDF are used, the initial Fayans functional
DF3 and its two variations, DF3-a and DF3-b, with different values
of spin-orbit parameters. Comparison is made with predictions of the
Skyrme--Hartree--Fock method with the HFB-17 functional. For the
DF3-a functional, phonon coupling (PC) corrections to
single-particle energies are found self-consistently with an
approximate account for the tadpole diagram. Accounting for the PC
corrections improves the agreement with the data for heavy nuclei, e.g.
for $^{208}$Pb. On the other hand, for lighter nuclei, e.g.,
$^{40,48}$Ca, PC corrections make the agreement a little worse. As
estimations show,  the main reason is that the approximation we use
for the tadpole term is less accurate for light nuclei.

\end{abstract}

\maketitle

\section{Introduction}
In the seminal article \cite{HF-VB} on the Hartree--Fock (HF)
method  with effective forces, Vautherin and  Brink   reduced
the effective Skyrme forces  containing a three-body term to much 
simpler version with a density dependent two-body force. Initially,
this dependence was assumed to be linear, just as that of the scalar
Landau--Migdal interaction amplitude in the theory of finite
Fermi systems (TFFS) \cite{AB1} playing the role of the effective
interaction in this approach. Inclusion of a velocity-dependent
force is another essential feature of the Skyrme HF (SHF) method. As
a result, the SHF effective Hamiltonian ${\cal H}_{\rm SHF}$ will
involve, in addition to the neutron and proton densities $
\rho_{n,p}({\bf r})$,  the kinetic-energy densities $\tau_{n,p}({\bf
r})$. The coordinate-dependent effective masses $m^*_{n,p}(r)$, as a
rule, differ significantly from the free-nucleon mass $m$. At
first sight,  this structure of the effective Hamiltonian seems to
contradict the Hohenberg--Kohn theorem \cite{Hoh-K}, which states
that the ground state energy of a Fermi system $E_0$ is a functional
of the density $ \rho({\bf r})$. However, as shown, {\it e.g.},
in \cite{Ring-Sch}, the kinetic energy $ \tau({\bf r})$ can be
expressed in terms of the density $\rho({\bf r})$, although the
relation is rather complicated.

Due to its simplicity, the SHF method quickly became very popular  and
up to now it dominates the self-consistent description of nuclear
properties. From the very beginning, the SHF method was aimed at
calculating global properties of nuclei, such as the binding energy
and average radii. There are numerous  sets of Skyrme force
parameters, some of them resulting in the description of nuclear masses
with a high accuracy.  The set HFB-17 \cite{HFB-17} led to a record
accuracy which is better, on average, than 600 keV.  We compare
our results for single-particle spectra we analyze with those
obtained with the HFB-17 functional.

At the same time, from the very beginning, the SHF method turned out
to be unsuccessful in describing single-particle spectra produced
by SHF mean-field potentials. The reason was the significant
deviation of the effective masses $m^*_{n,p}(r=0)\simeq 0.6\div0.8
m$ from the bare one typical for the SHF approach. In fact,  the
simplest shell model with Saxon--Woods potentials and $m^*=m$ was,
as a rule, more successful at this point. It is noteworthy that
the inclusion of single-particle energies to the fit of the SHF
parameters \cite{Brown} led to an effective mass close to the
bare one.

A bit later the self-consistent TFFS was developed. 
It  was based on the basic principles of the TFFS
\cite{AB1} supplemented with the condition of self-consistency
in the TFFS among the energy-dependent mass operator $\Sigma({\bf
r_1},{\bf r_2};\eps)$, the single-particle Green function $G({\bf
r_1},{\bf r_2};\eps)$, and the effective nucleon-nucleon (NN) interaction ${\cal
U}({\bf r_1},{\bf r_2},{\bf r_3},{\bf r_4};\eps,\eps')$
\cite{Fay-Khod}. The final version of this approach
\cite{Sap-Kh,KhS,AB2} was formulated in terms of the quasiparticle
Lagrangian ${\cal L}_q$, which is constructed to produce the
quasiparticle mass operator $\Sigma_q({\bf r},k^2;\eps)$. By
definition, the latter coincides at the Fermi surface with the exact
mass operator $\Sigma({\bf r},k^2;\eps)$. In the mixed
coordinate-momentum representation it depends linearly on the
momentum squared $k^2$ and the energy $\eps$ as well \cite{AB1}. In
magic nuclei which are nonsuperfluid, the Lagrangian ${\cal L}_q$
depends on three sorts of densities $\nu_i({\bf r}),i=0,1,2$ . The
first two densities are analogs of the SHF densities $\rho({\bf
r})$ and $\tau({\bf r})$, whereas the density $\nu_2({\bf r})$ is a new
ingredient of the self-consistent theory. It is the density of
single-particle energies which appears naturally due to the
$\eps$-dependence of the quasiparticle mass operator and determines
the $Z$-factor \beq Z({\bf r})=\frac 1 {1- \left(\frac {\partial
\Sigma} {\partial \eps}\right)_0}, \label{Z-fac}\eeq where the index
0 means that the energy and momentum  variables are taken at the
Fermi surface.

The self-consistent TFFS permits up to obtain the same bulk nuclear
characteristics as the SHF method. In addition, it helps to find the
$Z$-factor, which determines the in-volume component of the
one-nucleon $S$-factors. On equal footing, the TFFS from the very
beginning was focused on the analysis of the single-particle
spectra. The effective mass appearing in this approach contains not
only the so-called $k$-mass, as in the SHF method, but also the
``$E$-mass'': \beq \frac m {m^*({\bf r})}= Z({\bf
r})\left[1+2m \left(\frac {\partial \Sigma} {\partial k^2}\right)_0
\right]. \label{mstar} \eeq As found in \cite{KhS}, these two
ingredients of the effective mass should strongly  cancel each other
in order to describe the single-particle spectra of magic nuclei.
The optimal set of parameters found in \cite{KhS} corresponds to
the following characteristics of nuclear matter: $Z_0=0.8,\,
m^*_n=0.95,\, m^*_p=1.05$, which explains the success of the shell
model with $m^*=m$. For nuclear matter, a strong cancellation of the
$k$-mass and $E$-mass is well known in the Bruekner theory. It was
also analyzed  within the relativistic Bruekner--Hartree--Fock
method in Ref. \cite{Mahaux}.

It is noteworthy that corrections to the mean field theory due to
contributions of the low-lying surface vibrations, ``phonons'', were
involved in the analysis in \cite{KhS}.  All phonon coupling (PC)
diagrams were taken into account including so-called tadpole terms.
The method developed by Khodel \cite{Khodel} was used at that point.

Again, as in the SHF theory case, the appearance of a new density
$\nu_2({\bf r})$ does not contradict the Hohenberg--Kohn theorem.
As found in \cite{Kh-Sap-Zv}, it can be excluded if one goes
from the  quasiparticle Lagrangian ${\cal L}_q$ to the quasiparticle
Hamiltonian ${\cal H}_q$, which depends now on two densities, just
as the SHF Hamiltonian ${\cal H}_{\rm SHF}$. Moreover, if, on the
basis of the closeness  of the neutron and proton effective masses to
the bare one,  we put $m^*_n=m^*_p=m$, the Hamiltonian ${\cal H}_q$
will depend only on the  density $\rho({\bf r})$ normalized in a
standard way, just as in the energy density functional (EDF) of
Kohn--Sham \cite{K-Sh}. However, the quasiparticle Lagrangian  of
rather simple structure introduced in \cite{KhS} leads to a very
complicated density dependence of the Hamiltonian ${\cal
H}_q[\rho({\bf r})]$ \cite{Kh-Sap-Zv}   which could hardly be
introduced {\it ad hoc}.

The next important step in the self-consistent TFFS was made by
Fayans and coauthors \cite{Fay1}. On the base of the analysis in
\cite{Kh-Sap-Zv}, they formulated the theory directly in terms of
the EDF approach. They generalized the Kohn--Sham method to
superfluid systems, proposing for the normal component of the EDF the
fractional density dependence, with finite-range force, \beq
E_0=\int C_0 a\,f(|{\bf r}-{\bf r}'|) \frac {\rho({\bf r}') ^2} 2
\frac {1 - h_1 (\rho({\bf r}')/\rho_0)^{\alpha}} {1 + h_2 \rho({\bf
r}')/\rho_0}\, d^3 r\, d^3 r', \label{E0}\eeq where the factor
$C_0=(dn/d\eps_{\rm F})^{-1}$
 is the usual TFFS normalization factor, the inverse
density of states at the Fermi surface, and $\rho_0$ is the nuclear
matter density. The constants $a,h_1,h_2,\rho_0$, and $\alpha$ are
parameters and the Yukawa form for the finite range function $f(r)$
was used. Isotopic indices in (\ref{E0}) are omitted for brevity. In
Eq. (\ref{E0}), the spin-orbit and Coulomb interaction for protons
are omitted as well. For nuclear matter, the EDF (\ref{E0}), with
parameter values of \cite{Fay1} turned out to be very close to that
in \cite{Kh-Sap-Zv}. The identity $m^*=m$, which is a usual feature
of the Kohn--Sham method, was proposed in this approach. The explicit
form of the Fayans EDF and its different parametrizations DF1-DF3
can be found in \cite{Fay3,Fay} or, \cite{BE2}.

Recently, new data on single-particle spectra appeared
\cite{exp} for seven magic nuclei, from $^{40}$Ca to $^{208}$Pb. For
two of them, $^{78}$Ni and $^{100}$Sn, the spectra were not measured
directly but were interpolated from the  neighboring nuclei. The
bulk of these data contains 35 spin-orbit energy differences, which can be
used for fitting the spin-orbit and effective tensor force
parameters. In this article we carry out a comparative analysis of
these spectra within the EDF approach of Fayans {\it et al.} and the
SHF method with the set HFB-17 \cite{HFB-17}.

In addition, we analyze the PC corrections to single-particle
spectra including the tadpole term. The particle-vibration coupling
was extensively studied within  the so-called quasiparticle-phonon
model of Soloviev \cite{Solov} and within the ``nuclear-field'' approach
of Bortignon and Broglia \cite{Brog}. The use of phenomenological
parameters for single-particle spectra and particle-phonon coupling
constants was typical for these approaches. Evidently, the first
self-consistent consideration of the PC corrections to the
single-particle spectra was made by V. Bernard and Nguyen van Giai
\cite{Giai} within the SHF method. However, for a long time this
approach has been abandoned. Recently the interest in this problem
has been renewed. Self-consistent calculations with the SHF
functionals have been carried out in \cite{Vor} within the
quasiparticle-phonon model and in \cite{milan2,milan3,Bort} within
the nuclear-field method.  In a recent article \cite{Litv-Ring}
this problem was attacked within the relativistic mean field (RMF)
theory (see \cite{RMF}, and references therein).

Within the TFFS, the problem of PC corrections to $\eps_{\lambda}$
was examined in very old articles \cite{KhS,Platon}. An important
feature of these calculations was accounting for so-called tadpole
diagrams, which are ignored in all the approaches mentioned above. The
method developed by Khodel \cite{Khodel} is used for this aim.
However, these calculations were not completely self-consistent.
They used the Saxon--Woods basis, and the TFFS self-consistency
relation \cite{Fay-Khod} was taken into account approximately. In
this article, we follow the approach of \cite{KhS} and \cite{Platon}, enabling
complete self-consistency ,i.e., with the self-consistent basis and
self-consistent finding of the PC vertices $g_L$ for each of the
$L$-phonons. In addition, a wider number of magic nuclei is
considered for which single-particle spectra are available.

\section{EDF description of single-particle levels}

\begin{figure}[tbp]
\vspace{10mm} \centerline {\includegraphics [width=80mm]{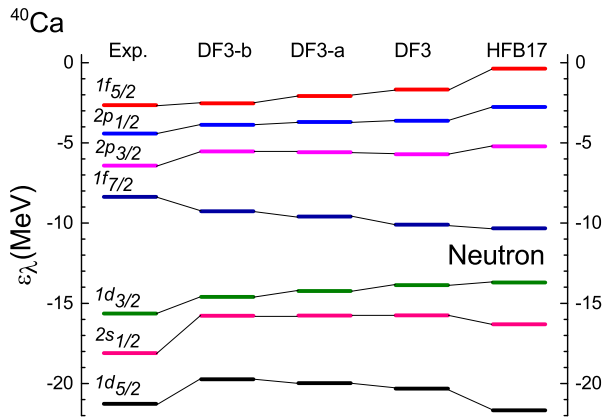}}
\vspace{2mm} \caption{ Neutron single-particle levels
in $^{40}$Ca. Experimental data from \cite{exp}.}
\end{figure}

\begin{figure}[tbp]
\vspace{10mm} \centerline {\includegraphics [width=80mm]{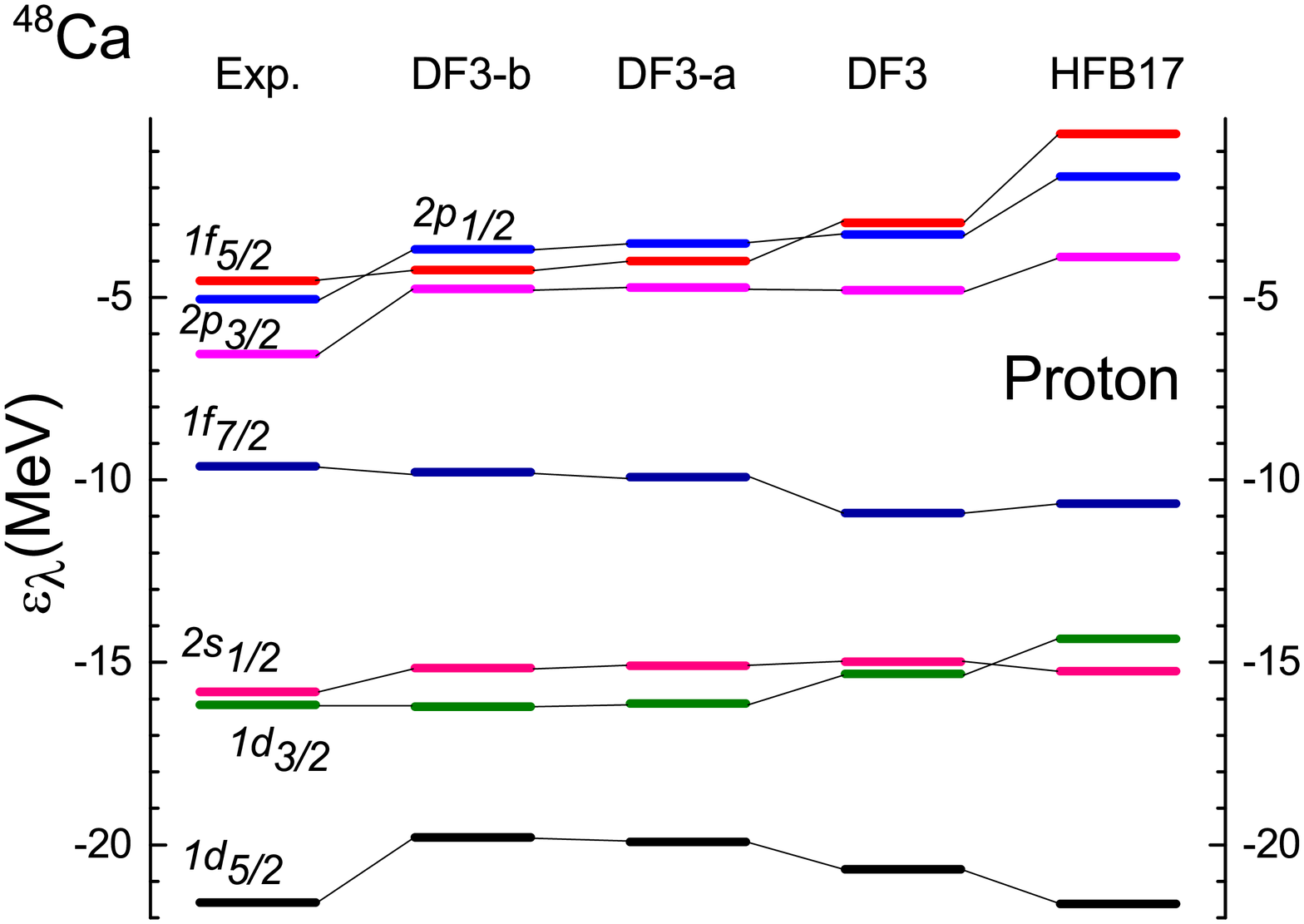}}
\vspace{2mm} \caption{ Proton single-particle levels
in $^{40}$Ca. Experimental data from \cite{exp}.}
\end{figure}

\begin{figure}[tbp]
\vspace{10mm} \centerline {\includegraphics [width=80mm]{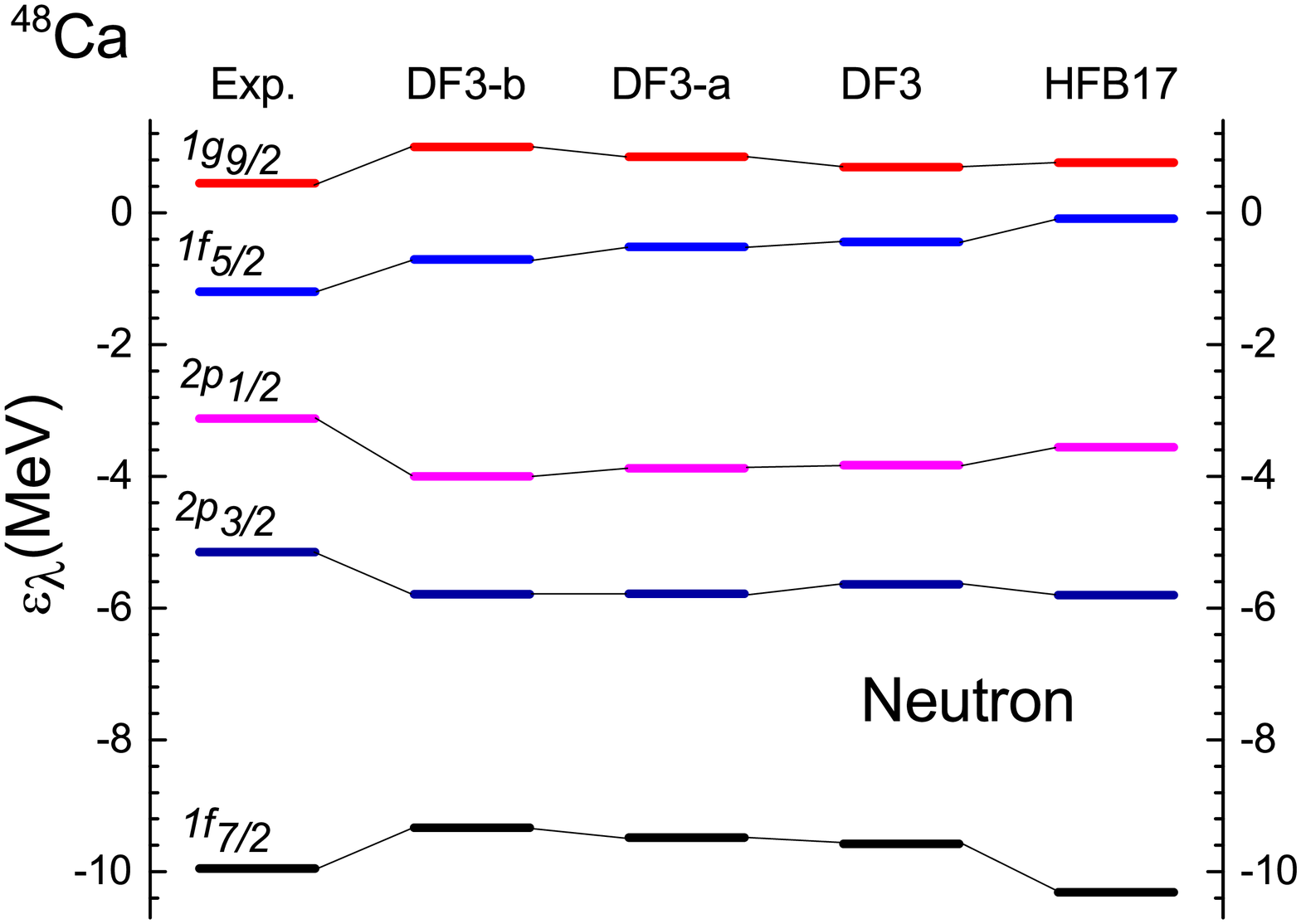}}
\vspace{2mm} \caption{ Neutron single-particle levels
in $^{48}$Ca. Experimental data from \cite{exp}.}
\end{figure}

\begin{figure}[tbp]
\vspace{10mm} \centerline {\includegraphics [width=80mm]{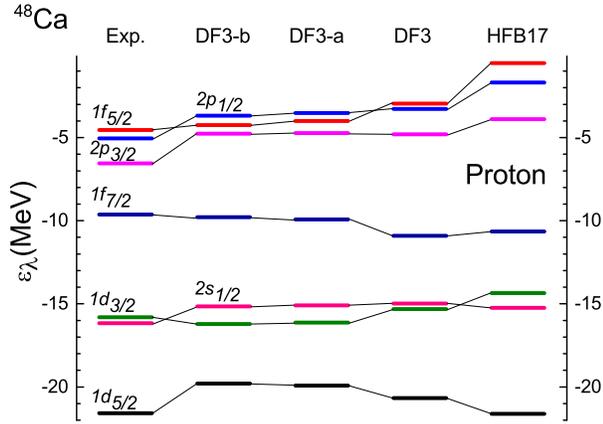}}
\vspace{2mm} \caption{ Proton single-particle levels
in $^{48}$Ca. Experimental data from \cite{exp}.}
\end{figure}

The parameter set DF3 \cite{Fay3} was used in the vest-known
application of the generalized EDF method of Fayans {\it et al.}
\cite{Fay}. This set not only was fitted to characteristics of
stable spherical nuclei from calcium to lead but also was specially fitted to
single-particle levels of the very neutron-rich doubly-magic nucleus
$^{132}$Sn. In Ref. \cite{Tol-Sap}, it was applied to nuclei of
uranium and transuranium regions which had not been analyzed previously
within this approach. It was found that for successful description
of this new bulk of nuclei, the spin-orbit parameters of the basic
DF3 set should be modified. To compare these two functionals
explicitly, we write down the spin-orbit terms of the EDF we
discuss.

\begin{figure}[tbp]
\vspace{10mm} \centerline {\includegraphics [width=80mm]{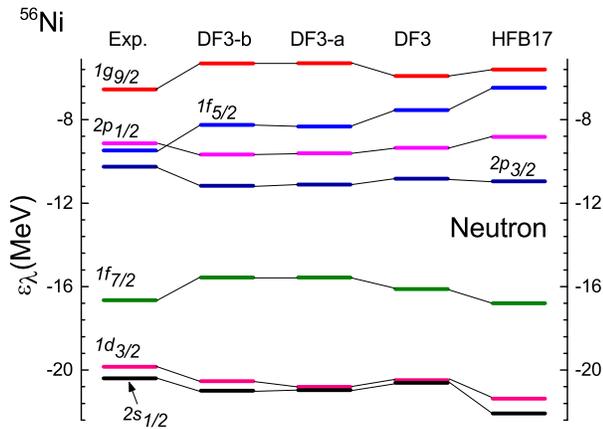}}
\vspace{2mm} \caption{ Neutron single-particle levels
in $^{56}$Ni. Experimental data from \cite{exp}.}
\end{figure}

\begin{figure}[tbp]
\vspace{10mm} \centerline {\includegraphics [width=80mm]{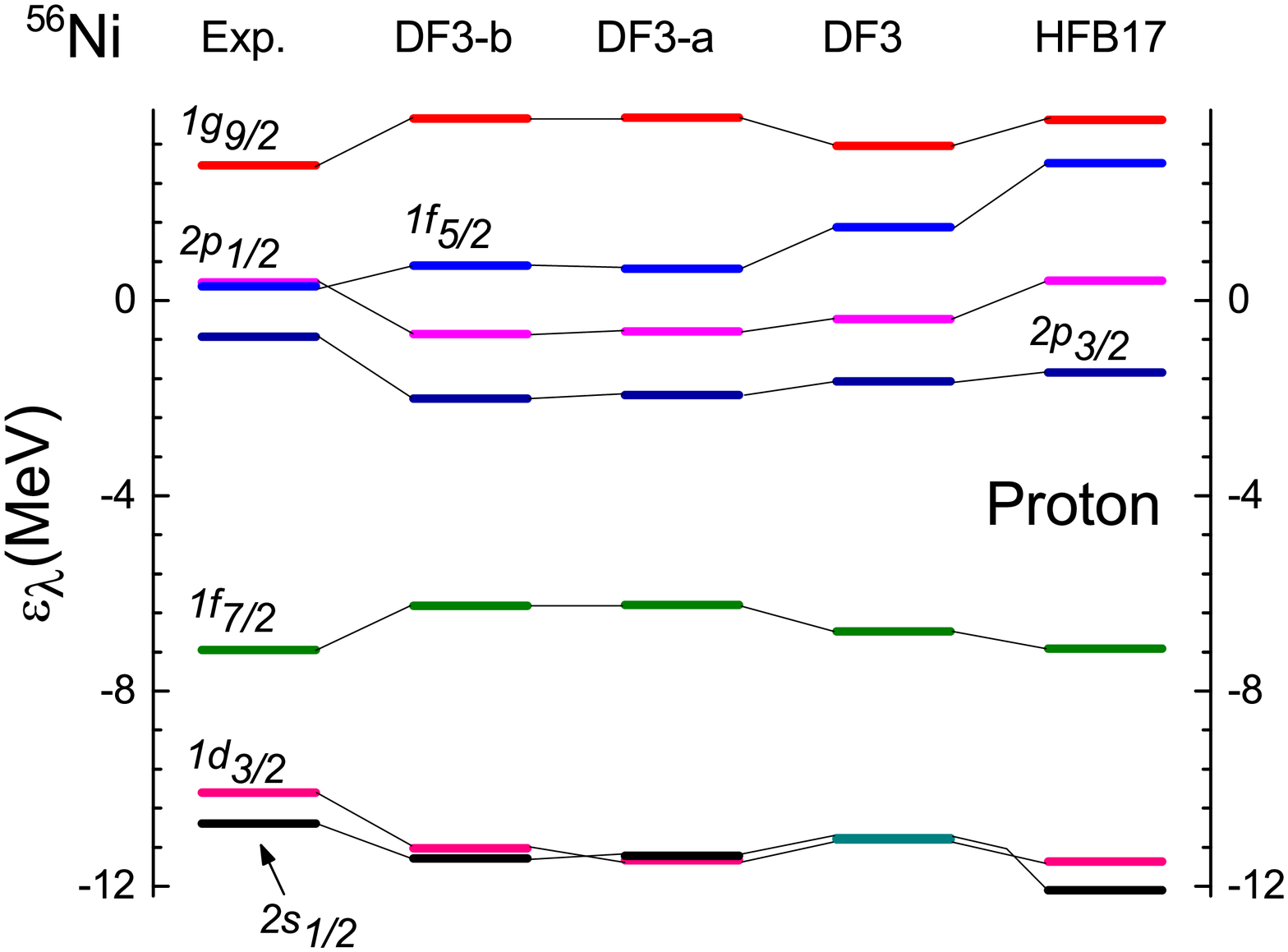}}
\vspace{2mm} \caption{ Proton single-particle levels
in $^{56}$Ni. Experimental data from \cite{exp}.}
\end{figure}

\begin{figure}[tbp]
\vspace{10mm} \centerline {\includegraphics [width=80mm]{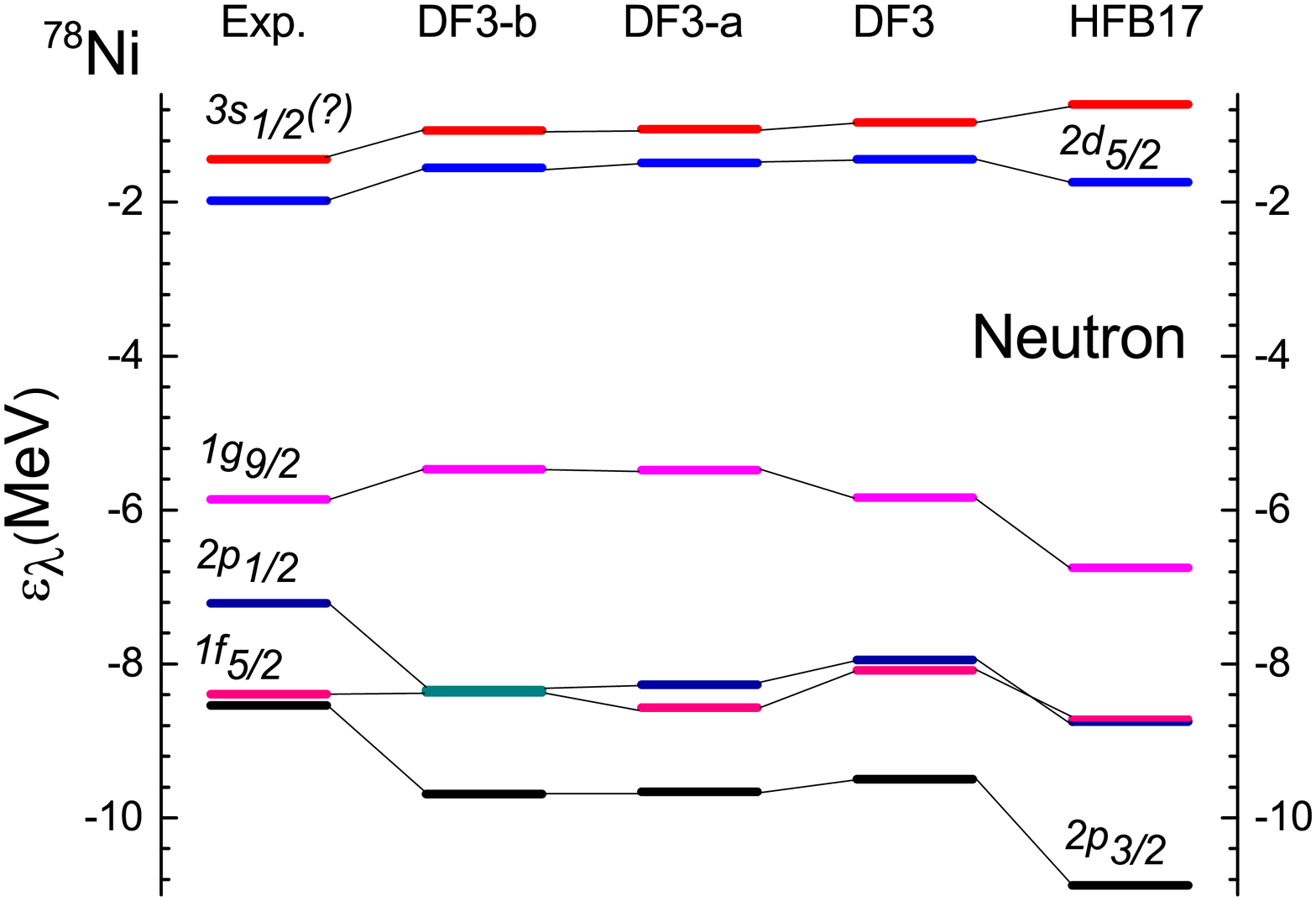}}
\vspace{2mm} \caption{ Neutron single-particle levels
in $^{78}$Ni. Experimental values \cite{exp} are interpolated from
data for neighboring nuclei.}
\end{figure}

\begin{figure}[tbp]
\vspace{10mm} \centerline {\includegraphics [width=80mm]{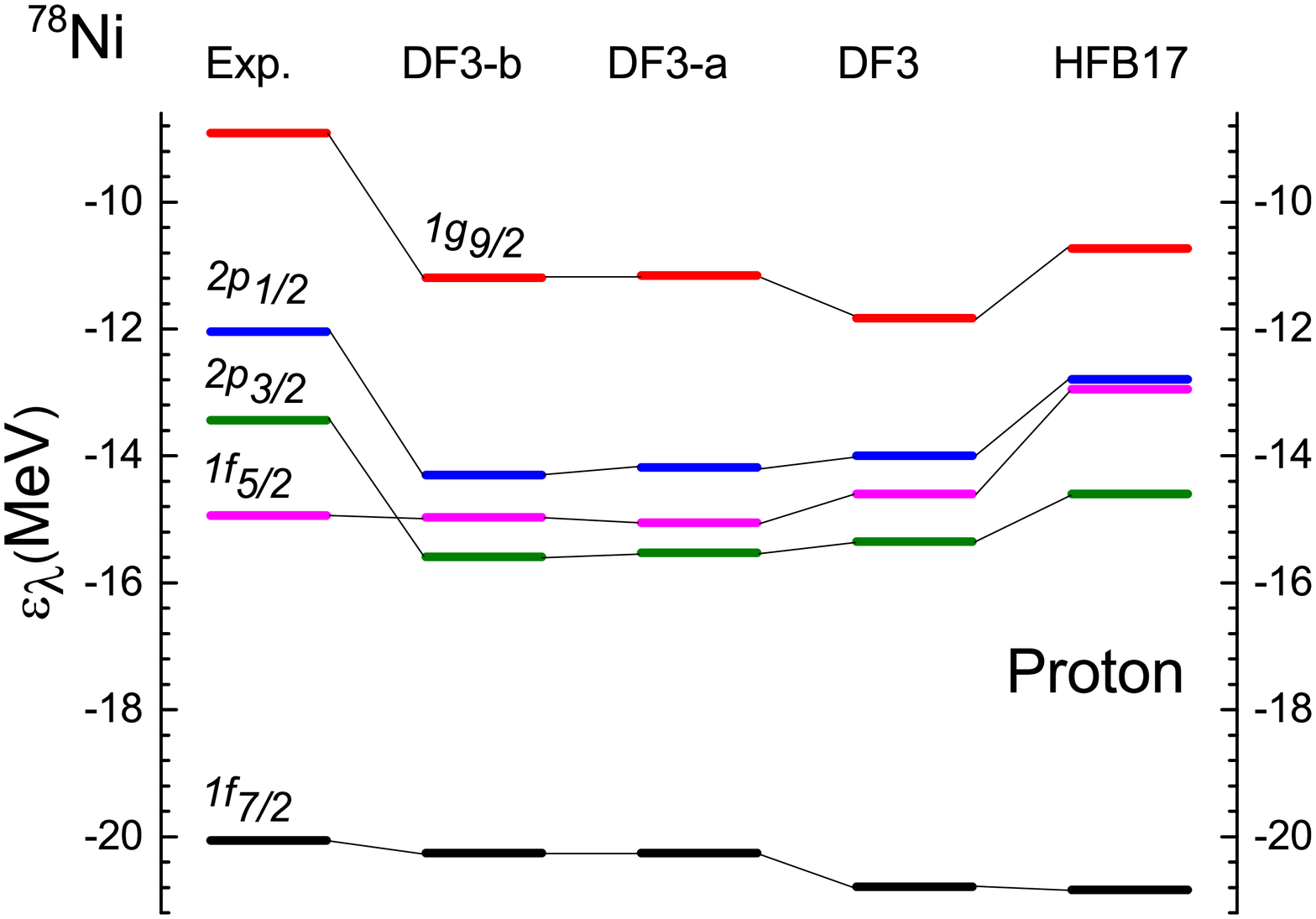}}
\vspace{2mm} \caption{ Proton single-particle levels
in $^{78}$Ni. Experimental values \cite{exp} are interpolated  from
data for neighboring nuclei.}
\end{figure}

The main spin-orbit effective interaction is taken in
\cite{Fay3,Fay} in the usual TFFS form, \beq
 {\cal F}_{sl}=C_0 r_0^2 (\kappa + \kappa' {\bfg\tau}_1{\bfg\tau}_2)
 \left[ \nabla_1 \delta ({\bf r}_1-{\bf r}_2) \times
 ({\bf p}_1-{\bf p}_2)\right]\cdot
({\bfg\sigma}_1 + {\bfg\sigma}_2), \label{Fsl} \eeq with obvious
notation.   Here the factor $r_0^2$ is introduced to make the
spin-orbit parameters  $\kappa$ and $\kappa'$ dimensionless. It can be
expressed in terms of the  equilibrium density $\rho_0$ of nuclear
matter introduced above,
 $r_0^2=(3/(8\pi \rho_0))^{2/3}$.

\begin{figure}[tbp]
\vspace{10mm} \centerline {\includegraphics [width=80mm]{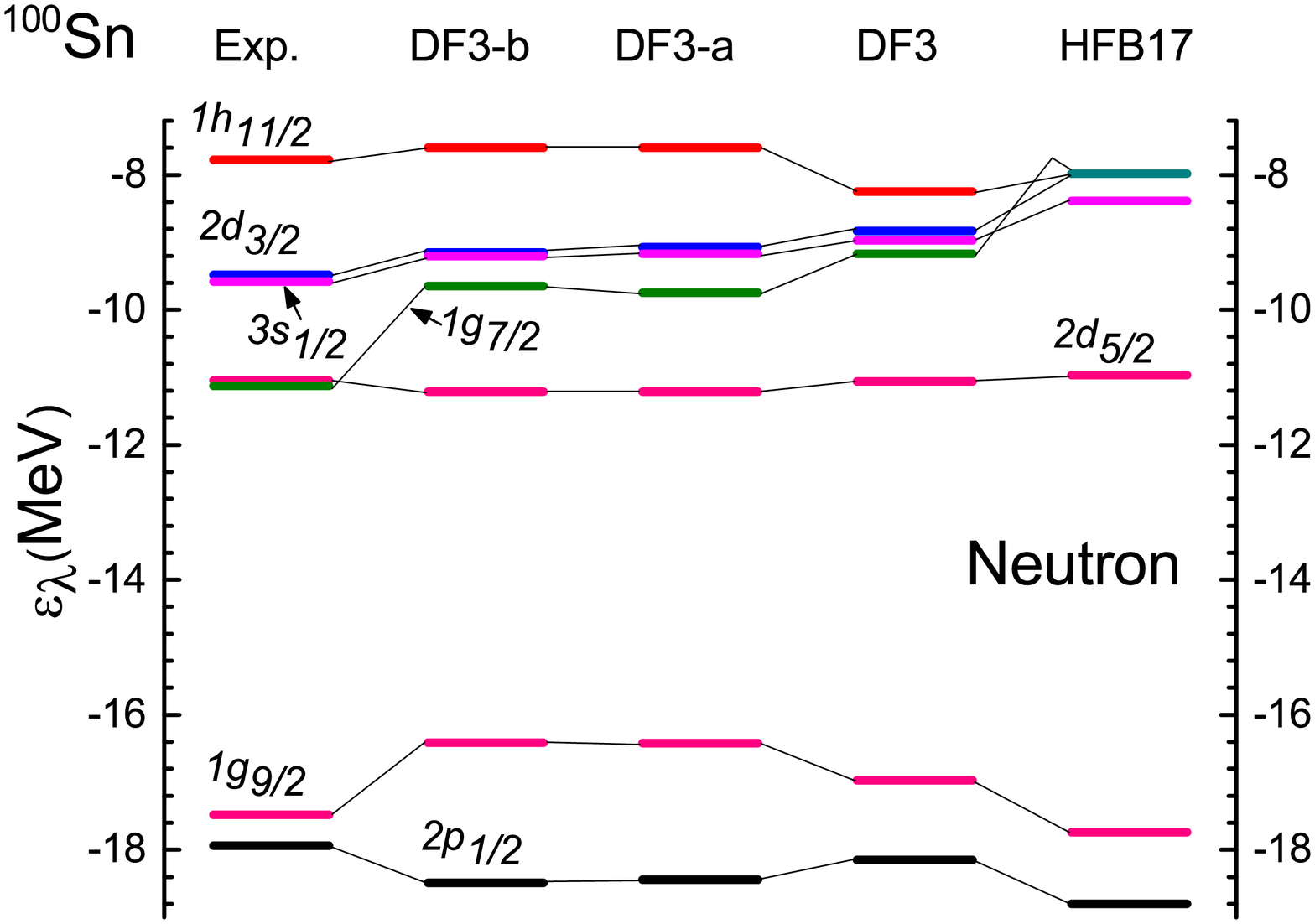}}
\vspace{2mm} \caption{ Neutron single-particle levels
in $^{100}$Sn. Experimental values \cite{exp} are interpolated  from
data for neighboring nuclei.}
\end{figure}

\begin{figure}[tbp]
\vspace{10mm} \centerline {\includegraphics [width=80mm]{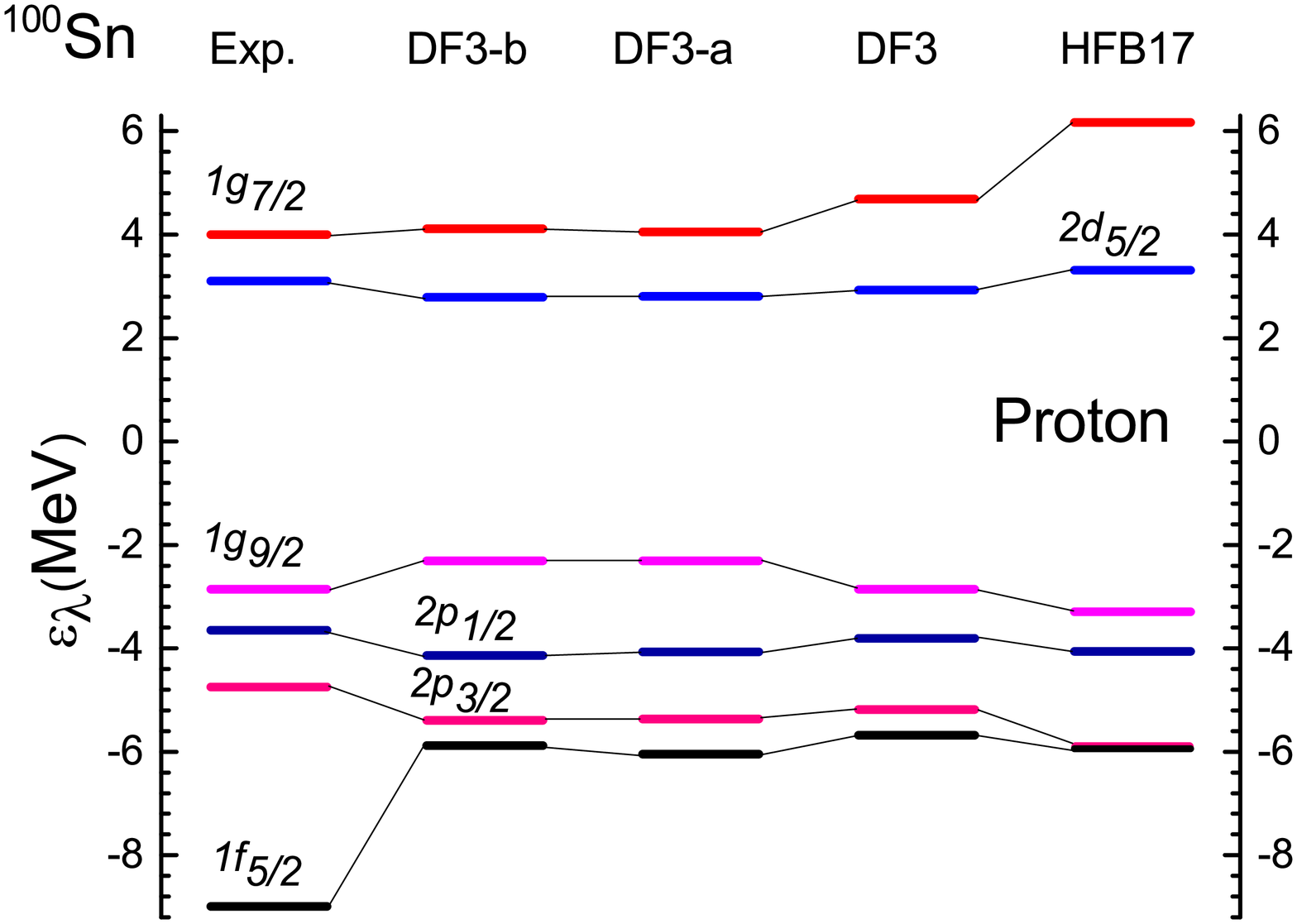}}
\vspace{2mm} \caption{ Proton single-particle levels
in $^{100}$Sn. Experimental values \cite{exp} are interpolated  from
data for neighboring nuclei.}
\end{figure}

\begin{figure}[tbp]
\vspace{10mm} \centerline {\includegraphics [width=80mm]{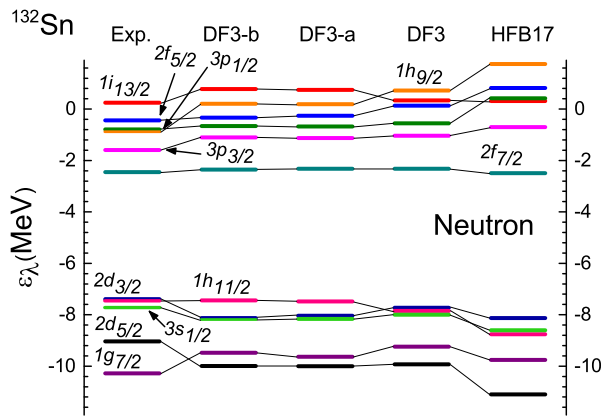}}
\vspace{2mm} \caption{ Neutron single-particle levels
in $^{132}$Sn. Experimental data from \cite{exp}.}
\end{figure}

\begin{figure}[tbp]
\vspace{10mm} \centerline {\includegraphics [width=80mm]{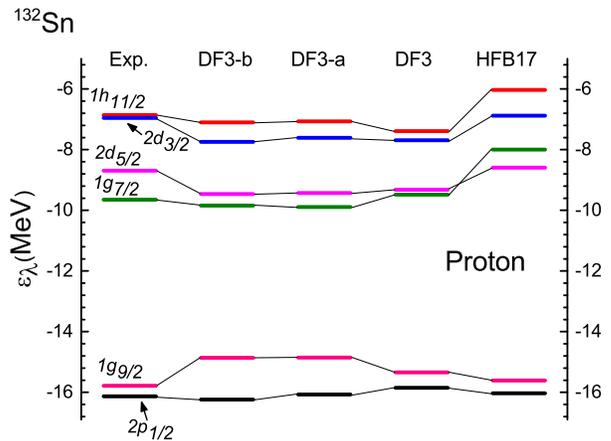}}
\vspace{2mm} \caption{ Proton single-particle levels
in $^{132}$Sn. Experimental data from \cite{exp}.}
\end{figure}

\begin{figure}[tbp]
\vspace{10mm} \centerline {\includegraphics [width=80mm]{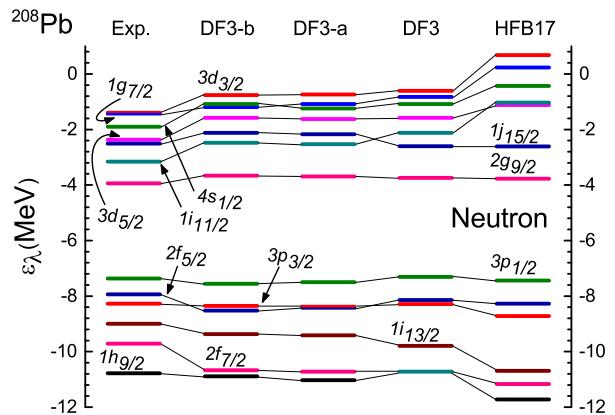}}
\vspace{2mm} \caption{ Neutron single-particle levels
in $^{208}$Pb. Experimental data from \cite{exp}.}
\end{figure}

\begin{figure}[tbp]
\vspace{10mm} \centerline {\includegraphics [width=80mm]{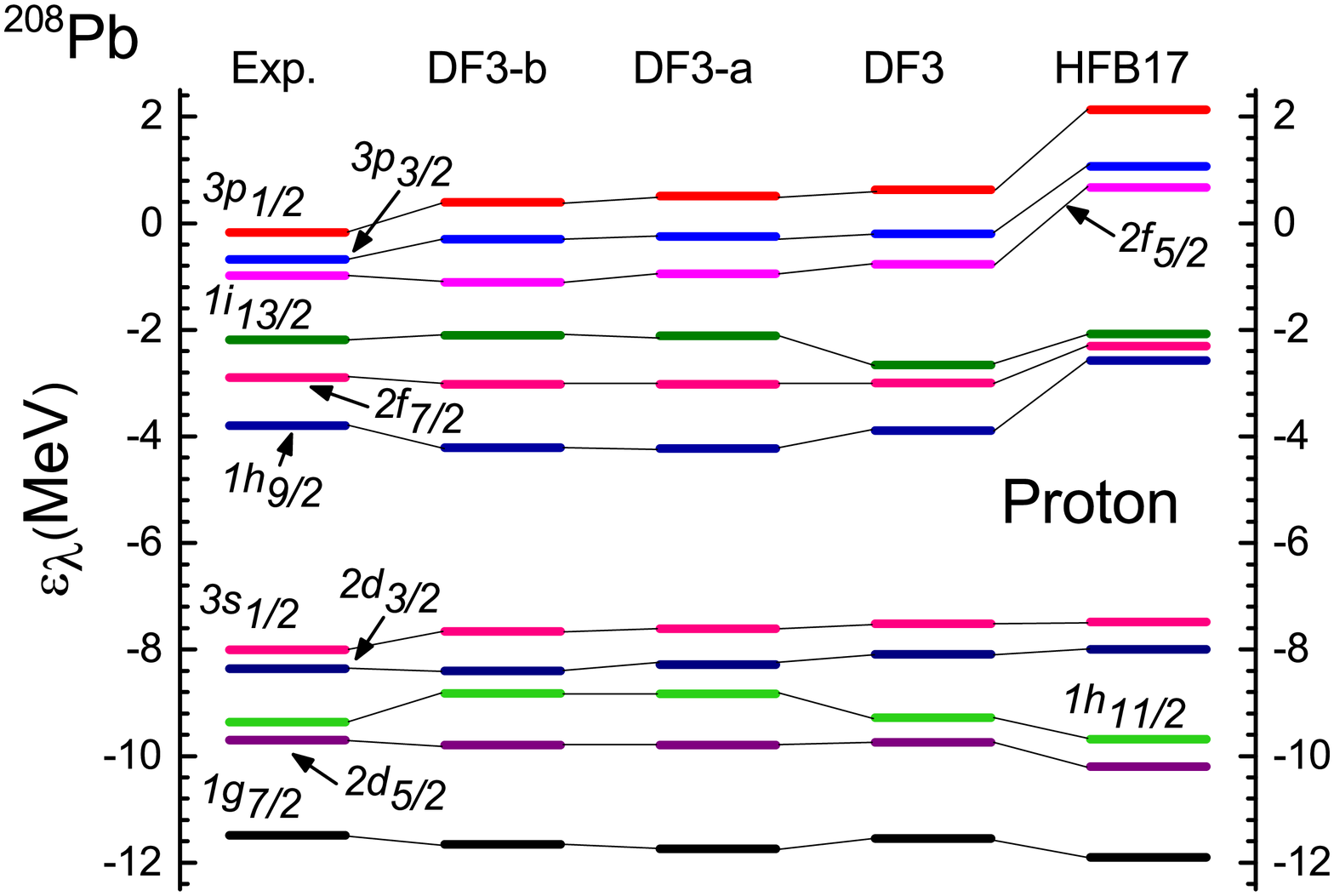}}
\vspace{2mm} \caption{ Proton single-particle levels
in $^{208}$Pb. Experimental data from \cite{exp}.}
\end{figure}

In nuclei with partially occupied spin-orbit doublets, the so-called
 spin-orbit density exists, \beq
\rho_{sl}^{\tau}(\bf r)=\sum_{\lambda} n_{\lambda}^{\tau} \langle
\phi_{\lambda}^{\tau *}(\bf r) ({\bfg \sigma}{\bf l})
\phi_{\lambda}^{\tau}(\bf r)\rangle, \label{rhosl} \eeq where $\tau
=n,p$  is the isotopic index and averaging over spin variables is carried out.
As is well known (see, e.g., \cite{KhS}), a new term appears in the
spin-orbit mean field induced by  the tensor forces and the first
harmonic $\hat{g_1}$ of the spin Landau--Migdal amplitude. We
combine those contributions into an effective tensor force or first
spin harmonic: \beq {\cal F}_1^s=C_0r_0^2 (g_1 + g'_1
{\bfg\tau}_1{\bfg\tau}_2)
 \delta ({\bf r}_1-{\bf r}_2) ({\bfg\sigma}_1 {\bfg\sigma}_2)
 ({\bf p}_1{\bf p}_2). \label{g-1}
\eeq For brevity, we call all four parameters in Eqs. (\ref{Fsl}) and (\ref{g-1})
spin-orbit parameters. 

\begin{table}
\caption{Spin-orbit parameters  of different versions of the Fayans
EDF.}
\begin{tabular}{c c c c }
\noalign{\smallskip}\hline\noalign{\smallskip}
Parameter & DF3 \cite{Fay}& DF3-a \cite{Tol-Sap}& DF3-b \\
\noalign{\smallskip}\hline\noalign{\smallskip}

$\kappa$      &  0.216     & 0.190  & 0.165 \\
$\kappa'$     &  0.077     & 0.077  & 0.075\\
$g_1$         &  0         & 0      & -0.100 \\
$g_1'$        & -0.123     & -0.308 & -0.300\\
\noalign{\smallskip}\hline\noalign{\smallskip}
\end{tabular}
\label{tab:param}
\end{table}

\begin{table} \caption{Deviations $\delta \Delta_{nls}$ (MeV) of the theory predictions
$\Delta_{nls}^{\rm theor}$   for  spin-orbit differences  from
experimental values for different functionals.\label{tab:ls}}
\begin{center}
\begin{tabular}{ccccccc}
\hline\noalign{\smallskip}

Nucleus & $\lambda$  & $\Delta_{nls}^{\rm exp}$  &  &
$\Delta_{nls}^{\rm theor}$ - &$\Delta_{nls}^{\rm exp}$ & \\
\noalign{\smallskip}
        &            &          & DF3-b &DF3-a &DF3& HFB17\\

\hline\noalign{\smallskip}
$^{40}$Ca-p & 1f & 5.69   &   0.69  &1.43 &   2.29  &      3.57\\
            & 1d & 5.40   &  -0.40  &0.22 &   0.90  &      2.35\\
            & 2p & 1.75   &  -0.35  &-0.16&   0.02  &      0.26\\

$^{40}$Ca-n & 1f & 5.71   &   0.99  &1.80 &   2.71  &      4.24\\
            & 1d & 5.63   &   -0.50 &0.12 &   0.82  &      2.34\\
            & 2p & 2.00   &   -0.34 &-0.12&   0.09  &      0.46\\

$^{48}$Ca-p & 1f & 5.08   &   0.46  &0.84  &  2.87  &      5.05\\
            & 1d & 5.77   &  -2.18  &-1.98 & -0.42  &      1.50\\
            & 2p & 1.50   &  -0.41  &-0.29 &  0.03  &      0.70\\
$^{48}$Ca-n & 1f & 8.75   &  -0.13  &0.21  &  0.39  &      1.47\\
            & 2p & 2.03   &  -0.24  &-0.13 & -0.22  &      0.21\\

 \noalign{\smallskip}\hline\noalign{\smallskip}

$^{56}$Ni-p & 1f & 7.45   &  -0.49  &-0.56 & 0.83  &      2.49\\
            & 2p & 1.11   &   0.21  &0.20 &  0.17  &      0.78\\
$^{56}$Ni-n & 1f & 7.17   &   0.14  &0.06 &  1.41  &      3.16\\
            & 2p & 1.11   &   0.39  &0.39 &  0.37  &      1.02\\

$^{78}$Ni-p & 1f & 5.12   &   0.17  &0.09 &  1.07  &      2.77 \\
            & 2p & 1.40   &  -0.11  &-0.05 &-0.05  &      0.41\\
$^{78}$Ni-n & 2p & 1.33   &   0.02  &0.06 &  0.22  &      0.80\\

 \noalign{\smallskip}\hline\noalign{\smallskip}

$^{100}$Sn-p& 1g & 6.86   &  -0.44  &-0.50 & 0.69  &      2.60\\
            & 2p & 1.10   &   0.15  &0.20 &  0.27  &      0.74\\
$^{100}$Sn-n& 1g & 6.35   &   0.41  &0.32 &  1.45  &      3.42\\
            & 2d & 1.57   &   0.49  &0.57 &  0.66  &      1.41\\

$^{132}$Sn-p& 1g & 6.13   &  -1.11  &-1.17 & -0.27 &      1.48\\
            & 2d & 1.74   &  -0.02  &0.08 &  0.19  &      0.82\\
$^{132}$Sn-n& 1h & 6.75   &   0.90  &0.92 &  1.82  &      3.76\\
            & 2f & 2.01   &   0.01  &0.05 &  0.44  &      1.30\\
            & 3p & 0.80   &  -0.36  &-0.35 & -0.31 &      0.32\\

\noalign{\smallskip}\hline\noalign{\smallskip}

$^{208}$Pb-p& 1h & 5.56   & -0.95   &-0.96 & -0.17 &      1.54\\
            & 2f & 1.92   & -0.01   &0.15  & 0.31  &      1.06\\
            & 2d & 1.34   &  0.05   &0.17  & 0.31  &      0.86\\
            & 3p & 0.85   & -0.16   &-0.09 &-0.02  &      0.21\\
$^{208}$Pb-n& 1i & 5.84   &  1.05   &1.03  & 1.83  &      3.82\\
            & 2g & 2.49   & -0.02   &0.12  & 0.42  &      1.51\\
            & 2f & 1.77   &  0.38   &0.53  & 0.81  &      1.68\\
            & 3d & 0.97   & -0.15   &-0.09 & 0.00  &      0.83\\

\noalign{\smallskip}\hline\noalign{\smallskip}

 $ \langle\delta \Delta_{nls}\rangle_{\rm rms}$       &   &  & 0.60  & 0.68 & 1.04 & 2.16 \\

 \noalign{\smallskip}\hline

\end{tabular}
\end{center}
\end{table}

The spin-orbit parameters of the set in \cite{Tol-Sap}, called DF3-a,
are listed in Table \ref{tab:param}, together with the initial set
DF3. Also, a new set, DF3-b, has been found for optimal description
of the spin-orbit energy differences. In Ref. \cite{exp} the bulk of
the data is given on spin-orbit doublets with known values of the
energies of both components, the total number being 35. This
provides us with the possibility of such optimization. The
experimental values of the spin-orbit differences,
$\Delta_{nls}=\eps_{n,l,j=l-1/2}-\eps_{n,l,j=l+1/2}$, for all magic
nuclei are listed in Table \ref{tab:ls} together with predictions
of the different functionals we analyze. For comparison with the SHF method,
we calculated also the single-particle spectra with the HFB-17
functional \cite{HFB-17}. To characterize the accuracy of all named
functionals in describing this specific set of data we found the
average theoretical error of predictions for each of them with the
expression \beq \label{deps_ls} \langle\delta
\Delta_{nls}\rangle_{\rm rms} = \sqrt{\frac 1 N \sum_{i=1}^N
\left(\Delta_{nls,i}^{\rm theor}-\Delta_{nls,i}^{\rm exp}\right)^2},
\eeq with obvious notation. The average error values are listed in
the last row in Table \ref{tab:ls}. Indeed,  the DF3-b version wins the
competition. The DF3-a functional describes the spin-orbit doublets
a little more poorly. For the DF3 version, the error increases to 1 MeV,
which is, however, twice as low as the HFB-17 result.

In Figs. 1--14. we compare the experimental data \cite{exp} with our
calculations employing three versions of the DF3 functional and using
the SHF functional HFB-17.
 To characterize the accuracy of a specific version, on average, we
calculated the corresponding average deviation of the theoretical
predictions from experiment for each magic nucleus, \beq
\label{deps_avr} \langle\delta \eps_{\lambda}\rangle_{\rm rms} =
\sqrt{\frac 1 N \sum_{\lambda} \left(\eps_{\lambda}^{\rm
theor}-\eps_{\lambda}^{\rm exp}\right)^2}; \eeq the summation
involves both neutrons and protons. The results are listed in Table
\ref{tab:error}. The row reports results of summation over
all nuclei. We see that the accuracy of all three versions of the
DF3 functional is significantly higher than that of the HFB-17
functional. We explain this with two important features of the Fayans
approach. First, it is the use of the bare mass $m^*=m$, which is
close to the prescription $m^*/m=1\pm 0.05$ of \cite{KhS}. Second,
the density dependence of the Fayans EDF (\ref{E0}) is essentially
more sophisticated than the SHF one. Being rather close to that in
\cite{Kh-Sap-Zv}, it involves implicitly the energy dependence of the
quasiparticle mass operator within the TFFS. Evidently, SHF
functionals turn out to be  oversimplified for describing successfully 
nuclear characteristics finer than the binding energies.

Among the three versions of the DF3 functionals,
the accuracy of the original one for spectra of magic nuclei is
a little higher. However,  the set DF3-a proved to be rather
successful, better than DF3, in describing characteristics of
semimagic nuclei such as the excitation energies and $B(E2)$ values of
the first $2^+$ states in even nuclei \cite{BE2,BE2-Web}  and
quadrupole moments of odd semimagic nuclei \cite{QEPJ,QEPJ-Web}. All
these quantities are very sensitive to the position of single-particle
levels in the vicinity of the Fermi surface. In addition, as 
mentioned above, the DF3-a functional works better for nuclei heavier
than lead. Therefore in the next section, dealing with PC
corrections to single-particle spectra, we use the DF3-a functional.

\begin{table}
\caption{Average deviations  $\langle\delta \eps_{\lambda}\rangle_{\rm rms}$ (MeV) of
the theory predictions for the single-particle energies from the
experimental values for magic nuclei. }
\begin{center}
\begin{tabular}{cccccc}
\noalign{\smallskip}\hline\noalign{\smallskip}

Nucleus   & $N$ &DF3-b & DF3-a &DF3& HFB17\\
\noalign{\smallskip}\hline\noalign{\smallskip}

$^{40}$Ca & 14& 1.08  & 1.25 & 1.35  & 1.64 \\

$^{48}$Ca & 12& 0.89   &1.00 & 1.01  & 1.70 \\

$^{56}$Ni & 14& 1.00   &0.97 & 0.85  & 1.40 \\

$^{78}$Ni & 11& 1.24   & 1.41 & 1.09  & 1.32 \\

$^{100}$Sn& 13& 1.09 & 1.17 & 1.01 & 1.56 \\

$^{132}$Sn& 17& 0.58  & 0.66 & 0.55 & 1.15  \\

$^{208}$Pb& 24& 0.44  & 0.51 &0.43 & 1.15 \\

\noalign{\smallskip}\hline\noalign{\smallskip}

Total & 105& 0.89  & 0.98 &0.89 & 1.40 \\

\noalign{\smallskip}\hline\noalign{\smallskip}
\end{tabular}
\end{center}
\label{tab:error}
\end{table}

\section{Phonon coupling corrections to single-particle energies}

Accounting for PC effects, the equation for single-particle
energies and wave functions  can be written as \beq \left(\eps-H_0
-\delta \Sigma^{\rm PC}(\eps) \right) \phi =0, \label{sp-eq}\eeq where
$H_0$ is the quasiparticle Hamiltonian with the spectrum
$\eps_{\lambda}^{(0)}$ and $\delta \Sigma^{\rm PC}$ is the PC
correction to the quasiparticle mass operator. After expanding this
term in the vicinity of $\eps=\eps_{\lambda}^{(0)}$ one finds \beq
\eps_{\lambda}=\eps_{\lambda}^{(0)} + Z_{\lambda}^{\rm PC} \delta
\Sigma^{\rm PC}_{\lambda\lambda}(\eps_{\lambda}^{(0)})
,\label{eps-PC}\eeq with obvious notation. Here $Z^{\rm PC}$ denotes
the $Z$-factor due to the PC effects, i.e. that found from Eq.
(\ref{Z-fac}) with substitution of $\delta \Sigma^{\rm PC}(\eps)$ instead
of the main mass operator $\Sigma(\eps)$. Remember that in the TFFS the
corresponding $Z$-factor is included in the quasiparticle
Hamiltonian $H_0$. For brevity, below the superscript PC is
omitted. Expression (\ref{eps-PC}) corresponds to the perturbation
theory in the $\delta \Sigma$ operator with respect to $H_0$. In this
article, we limit ourselves to magic nuclei where the so-called
$g_L^2$-approximation, $g_L$ being the $L$-phonon creation
amplitude, is, as a rule, valid. It is worth mentioning that Eq.
(\ref{eps-PC}) is more general, including, say, $g_L^4$ terms.

Let us now consider $g_L^2$-corrections to the quasiparticle mass
operator (Fig. 15).  The first, pole diagram is well examined and
corresponding equations can be found in textbooks, e.g., in
\cite{AB1,AB2}. Therefore we concentrate mainly on the second,
tadpole term which has not been as widely discussed in the
literature.

The vertex $g_L$  in Fig. 15 obeys the equation
\cite{AB1} \beq { g_L}(\omega)={{\cal F}} {A}(\omega) {
g_L}(\omega), \label{g_L} \eeq
 where $ A(\omega)=\int
G \left(\eps + \omega/ 2 \right) G \left(\eps - \omega/ 2 \right)d
\eps/(2 \pi i)$ is the particle-hole propagator, $G(\eps)$ being the
one-particle Green function. In obvious symbolic notation, the pole
diagram corresponds to  $\delta\Sigma^{\rm pole}=(g_L,DGg_L)$, where
$D_L(\omega)$ is the phonon $D$-function, or explicitly one obtains
\bea \delta\Sigma^{\rm
pole}_{\lambda\lambda}(\epsilon)&=&\sum_{\lambda_1\,M}
|\langle\lambda_1|g_{LM}|\lambda\rangle|^2 \nonumber\\
&\times&\left(\frac{n_{\lambda_1}}{\eps+\omega_L-
\eps_{\lambda_1}}+\frac{1-n_{\lambda_1}}{\eps-\omega_L
-\eps_{\lambda_1}}\right), \label{dSig2} \eea where $\omega_L$ is
the excitation energy of the $L$-phonon and $n_{\lambda}=(0,1)$
stands for the occupation numbers.

\begin{figure}[tbp]
\vspace{10mm} \centerline {\includegraphics [width=80mm]{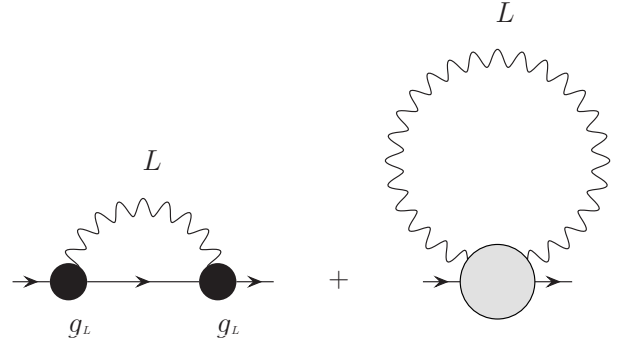}}
\vspace{2mm} \caption{PC corrections to the mass operator. The gray
circle denotes the ``tadpole'' term.}
\end{figure}

All the low-lying phonons we consider have  natural parity. In
this case, the vertex $g_L$ possesses  even $T$-parity. It is a
sum of two components with spins $S=0$ and $S=1$, respectively, \beq
g_L= g_{L0}(r) T_{LL0}({\bf n,\alpha}) +  g_{L1}(r)
T_{LL1}({\bf n,\alpha}), \label{gLS01} \eeq where $T_{JLS}$ stand
for the usual spin-angular tensor operators \cite{BM1}. The
operators $T_{LL0}$ and $T_{LL1}$ have  opposite $T$-parities, hence
the spin component should be the odd function of the excitation
energy, $g_{L1}\propto \omega_L$.
For the ghost dipole, $L=1$ and $\omega_1=0$, Eq. (\ref{g_L}), due
to the TFFS self-consistency relation \cite{Fay-Khod}, has the exact
solution \beq g_1 ({\bf r}) = \alpha_1 \frac {d U(r) } {d r}Y_{1M}({\bf n}), \label{g1}\eeq where $\alpha_1=
 1 /\sqrt{2\omega
B_1}$ , $B_1=3m A/4\pi $ is the Bohr--Mottelson (BM) mass coefficient \cite{BM2} and
$U(r)$ is the central part of the mean-field potential generated
by the energy functional.

For the ghost phonon  it is convenient to rewrite Eq. (\ref{dSig2})
as follows: \bea \delta\Sigma^{\rm
pole}_{\lambda\lambda}(\epsilon)=\alpha_1^2\sum_{\lambda_1\,M}
\left|\langle\lambda_1| \frac {dU}{dr} Y_{1M}|\lambda\rangle\right|^2 \nonumber\\
\times\left(\frac{\eps - \eps_{\lambda_1}}{(\eps -
\eps_{\lambda_1})^2 -\omega_1^2 }+ \omega_1\frac{1-2
n_{\lambda_1}}{(\eps - \eps_{\lambda_1})^2 -\omega_1^2 }\right).
\label{dSig3} \eea

The second, tadpole, term in Fig. 15  is \beq \delta\Sigma^{\rm
tad}=\int \frac {d\omega} {2\pi i} \delta_L {g_L}
D_L(\omega),\label{tad} \eeq where $\delta_L {g_L}$  can be found
\cite{Khodel,KhS} by variation of Eq. (\ref{g_L}) in the field of
the $L$-phonon: \bea  \delta_L {g_L}&=&\delta_L {\cal F}
A(\omega_L){ g_L} + {\cal F} \delta_
L A(\omega_L){ g_L} \nonumber \\
&+& {\cal F} A(\omega_L)\delta_L{g_L}. \label{dgL} \eea The phonon
$D$-function appears in Eq. (\ref{tad}) after connecting  two wavy
phonon ends in Eq. (\ref{dgL}). This corresponds to averaging of the
product of two boson (phonon) operators $B_L^+B_L$ over the ground
state of the nucleus with no phonons.

The quantity $\delta_L A$ can be readily obtained by variation of
each Green function in the particle-hole propagator $A$ in field
$g_L$ induced by the $L$-phonon. The explicit expression for the
variation $\delta_L {\cal F}$ can not be found within the TFFS as in
this approach the Landau--Migdal amplitude ${\cal F}$ is introduced as a
phenomenological quantity. In Ref. \cite{KhS} the ansatz was
proposed, \beq  \delta_L {\cal F} = \frac {\delta {\cal F}(\rho)}
{\delta \rho} \delta \rho_L,\label{dLF}\eeq where \beq \delta \rho_L
=A_L g_L   \label{rhoL} \eeq is the transition density for
excitation of the $L$-phonon.

The complete  PC correction from the $L$-phonon to the single
particle energy is \beq \delta \eps_{\lambda} =
Z_{\lambda}\left(\delta\Sigma^{\rm pole}_{\lambda\lambda} +
\delta\Sigma^{\rm tad}_{\lambda\lambda}\right). \label{deps} \eeq As
the term $\delta\Sigma^{\rm tad}$ does not depend on the energy
$\eps$, it does not contribute to $Z_{\lambda}$. Hence, the PC
contribution to the $Z$-factor is \beq Z_{\lambda}= \frac 1
{1-\left.\frac {\partial} {\partial\eps}
\delta\Sigma_{\lambda\lambda}^{\rm pole}(\eps)
\right|_{\eps=\eps_{\lambda}}}. \label{ZPC} \eeq The explicit
relation for energy derivative of the mass operator (\ref{dSig2})
can be easily  obtained.

\begin{figure}[tbp]
\vspace{10mm} \centerline {\includegraphics [width=80mm]{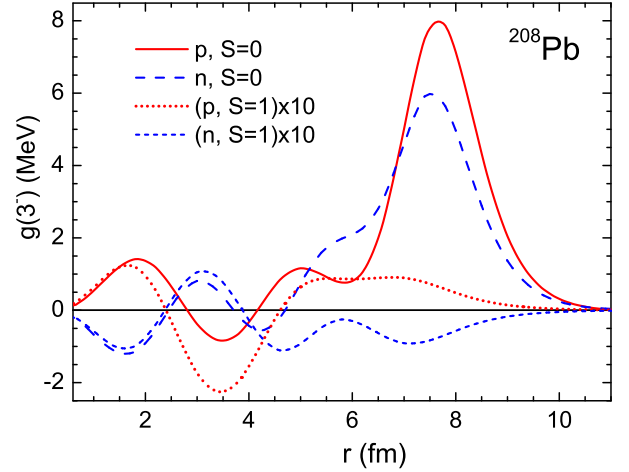}}
\vspace{2mm} \caption{ The vertex $g_L$ for the
$3^-_1$ state in $^{208}$Pb.} \label{fig:g3mi}\end{figure}

Dealing with the ghost phonon, Eqs. (\ref{tad}) and (\ref{dgL}),
with the use of (\ref{g1}),  can be transformed \cite{KhS} to  \beq
\delta\Sigma^{\rm tad}_{L=1} = \frac {\alpha_1^2} {2} \triangle U(r). \label{dg1}\eeq

For the ghost phonon both terms of the sum (\ref{deps}) are
proportional to $\alpha_1^2\propto 1/\omega_1$, hence the
$\omega_1$-even component of Eq. (\ref{dSig3})  and the tadpole term
(\ref{dg1}) should compensate each other: \bea
\alpha_1^2\sum_{\lambda_1\,M} \left|\langle\lambda_1| \frac {dU}{dr}
Y_{1M}({\bf n})|\lambda\rangle\right|^2 \frac{\eps_{\lambda} -
\eps_{\lambda_1}}{(\eps_{\lambda} - \eps_{\lambda_1})^2 -\omega_1^2
}\nonumber\\ + \left(\delta\Sigma^{\rm
tad}_{L=1}\right)_{\lambda\lambda}=0 . \qquad \qquad \qquad
\label{deps1} \eea This identity could be proved explicitly
\cite{KhS,Platon} with the use of the identity $(\partial U/
\partial {\bf r})_{\lambda\lambda'}=(\eps_{\lambda'} -
\eps_{\lambda})(\partial/
\partial {\bf r})_{\lambda\lambda'} $.

In the result one obtains \bea \delta \eps_{\lambda}^{\rm ghost} &=&
 \frac 1 {2B_1}\sum_{\lambda_1\,M}
\left|\langle\lambda_1| \frac {dU}{dr} Y_{1M}({\bf n})|\lambda\rangle\right|^2 \nonumber\\
& \times & \frac{1 - 2n_{\lambda_1}}{(\eps_{\lambda} -
\eps_{\lambda_1})^2 -\omega_1^2 }\,. \label{depsghoust}    \eea
The physical meaning of the PC correction caused by the ghost $1^-$-phonon is very
simple. This is account for the ``recoil effect'' due to the center-of-mass (CM) motion.
Equation (\ref{depsghoust}), with the use of the above relation for the $(\partial U/
\partial {\bf r})$ operator, can be reduced to the usual RPA formula for CM motion correction, \cite{Ring-Sch}:
\beq  \delta \eps_{\lambda}^{\rm ghost} =
 \frac 1 {2B_1}\sum_{\lambda_1\,M} |{\bf k}_{{\lambda}_1{\lambda}}|^2;   \label{ghoust1} \eeq
 although more cumbersome Eq. (\ref{depsghoust}) is convenient for numerical calculations.

\begin{table}[]
\caption{Characteristics of the low-lying phonons in magic nuclei,
$\omega_L$ (MeV) and $B(EL,{\rm up)}$(${\rm e^2 fm}^{2L}$).}
\begin{tabular}{c c c c c  }

\noalign{\smallskip}\hline\noalign{\smallskip} $L^\pi$  &
$\omega_L^{\rm th}$   & $\omega_L^{\rm exp}$
&  $B(EL)^{\rm th}$ &  $B(EL)^{\rm exp}$  \\
\noalign{\smallskip}\hline\noalign{\smallskip}

&&    $\qquad\qquad ^{40}$Ca          &&\\
\noalign{\smallskip}
$3^-$    & 3.335     &  3.73669 (5)       &  $1.52 \times 10^4$                  & $1.24   \times 10^4 $ \\
\noalign{\smallskip}\hline\noalign{\smallskip}
&&    $\qquad\qquad ^{48}$Ca          &&\\
\noalign{\smallskip}
$2^+$    & 3.576     &  3.83172 (6)       &  $0.55   \times 10^2 $    & $0.86   \times 10^2 $ \\

$3^-$    & 4.924     &  4.50678 (5)        &$5.701   \times 10^3 $    & $0.67   \times 10^4 $ \\
\noalign{\smallskip}\hline\noalign{\smallskip}
&&    $\qquad\qquad ^{56}$Ni          &&\\
\noalign{\smallskip}
$2^+$    & 2.826     &   2.7006 (7)        & $5.725 \times 10^2 $                   &  \\

$3^-$    & 8.108     &   4.932 (3)         &$2.068   \times 10^4 $    & \\
\noalign{\smallskip}\hline\noalign{\smallskip}
&&    $\qquad\qquad ^{78}$Ni          &&\\
\noalign{\smallskip}
$2^+$    & 3.238     & -         & $3.309 \times 10^2$      & \\

$3^-$    & 6.378     & -         &$1.549   \times 10^4 $    & \\
\noalign{\smallskip}\hline\noalign{\smallskip}
&&    $\qquad\qquad ^{100}$Sn          &&\\
\noalign{\smallskip}
$2^+$    & 3.978     & -    & $1.375\times 10^3$            & \\

$3^-$    & 5.621     & -         &$1.24   \times 10^5 $    & \\

\noalign{\smallskip}\hline\noalign{\smallskip}
&&    $\qquad\qquad ^{132}$Sn          &&\\
\noalign{\smallskip}
$2^+$    & 4.327      & 4.04120 (15)    & $0.104\times 10^4$    & $0.11 (0.03)\times 10^4$\\

$3^-$    & 4.572      & 4.35194 (14)         &$1.29   \times 10^5 $    & \\

\noalign{\smallskip}\hline\noalign{\smallskip}
&&    $\qquad\qquad ^{208}$Pb          &&\\
\noalign{\smallskip}
$3^-$      & 2.684    &  2.615       &$7.093\times 10^5 $    & $6.12 \times 10^5 $\\
$5^-_1$    & 3.353    &  3.198     &$3.003\times 10^8 $    & $4.47 \times 10^8 $\\
$5^-_2$    & 3.787    &  3.708     &$1.785 \times 10^8 $   & $2.41 \times 10^8 $\\
$2^+_1$    & 4.747    &  4.086     &$1.886 \times 10^3 $   & $3.18 \times 10^3 $\\
$2^+_2$    & 5.004    &  4.928     &$1.148 \times 10^3 $   & - \\
$4^+_1$    & 4.716    &  4.324     &$3.007 \times 10^6 $   & - \\
$4^+_2$    & 5.367    &  4.911(?)  &$8.462 \times 10^6 $   & - \\
$6^+_1$    & 4.735    &   -        &$6.082 \times 10^9 $   & - \\
$6^+_2$    & 5.429    &   -        &$1.744 \times 10^{10}$ & - \\

\noalign{\smallskip}\hline\noalign{\smallskip}

\end{tabular}\label{tab:BEL}
\end{table}

\begin{table}
\caption{Pole and tadpole contributions to PC corrections from
$3^-$-states to single-particle energies (MeV) in $^{40}$Ca. }
\begin{tabular}{cccc}
\noalign{\smallskip}\hline\noalign{\smallskip}

$\lambda$ &  $\delta \eps^{\rm pole}_{\lambda}$ & $\delta \eps^{\rm
tad}_{\lambda}$ & $\delta \eps_{\lambda}$ \\
\noalign{\smallskip}\hline\noalign{\smallskip} \noalign{\smallskip}
         &           neutr.     &        & \\
\noalign{\smallskip}
$1f_{5/2}$ & -0.395 &  0.592 &  0.197 \\
$2p_{1/2}$ & -0.805 &  0.305 & -0.500 \\
$2p_{3/2}$ & -0.833 &  0.383 & -0.450 \\
$1f_{7/2}$ & -0.142 &  0.733 &  0.591 \\
$1d_{3/2}$ & -0.426 &  0.697 &  0.271 \\
$2s_{1/2}$ & -0.932 &  0.493 & -0.439 \\
$1d_{5/2}$ & -0.253 &  0.731 &  0.478 \\

\noalign{\smallskip}\hline\noalign{\smallskip}
         &           prot.     &        & \\
\noalign{\smallskip}
$1f_{5/2}$ & -0.240 &  0.470 &  0.230 \\
$2p_{1/2}$ & -0.584 &  0.152 & -0.432 \\
$2p_{3/2}$ & -0.224 &  0.251 &  0.027 \\
$1f_{7/2}$ & 0.100  & 0.677  &  0.777 \\
$1d_{3/2}$ & -0.370 &  0.659 &  0.289 \\
$2s_{1/2}$ & -0.886 &  0.429 & -0.457 \\
$1d_{5/2}$ & -0.234 &  0.699 &  0.466 \\
\end{tabular}\label{tab:Ca-tad}
\end{table}

\begin{table}
\caption{Pole and tadpole contributions to PC corrections from
$3^-$-states to single-particle energies (MeV) in $^{208}$Pb. }
\begin{center}
\begin{tabular}{cccc}
\noalign{\smallskip}\hline\noalign{\smallskip}

$\lambda$ &  $\delta \eps^{\rm pole}_{\lambda}$ & $\delta \eps^{\rm
tad}_{\lambda}$ & $\delta \eps_{\lambda}$ \\
\noalign{\smallskip}\hline\noalign{\smallskip}
         &   neutr.            &            &            \\
\noalign{\smallskip}
$3d_{3/2}$ & -0.150 &  0.012 & -0.137 \\
$2g_{7/2}$ & -0.142 &  0.061 & -0.081 \\
$4s_{1/2}$ & -0.134 &  0.016 & -0.118 \\
$3d_{5/2}$ & -0.147 &  0.023 & -0.124 \\
$1j_{15/2}$& -0.708 &  0.204 & -0.504 \\
$1i_{11/2}$& -0.058 &  0.198 &  0.140 \\
$2g_{9/2}$ & -0.244 &  0.076 & -0.167 \\
$3p_{1/2}$ & -0.220 &  0.053 & -0.167 \\
$2f_{5/2}$ & -0.186 &  0.094 & -0.092 \\
$3p_{3/2}$ & -0.205 &  0.056 & -0.149 \\
$1i_{13/2}$&  0.057 &  0.211 &  0.269 \\
$2f_{7/2}$ &  0.724 &  0.091 &  0.815 \\
$1h_{9/2}$ & -0.014 &  0.197 &  0.184 \\

\noalign{\smallskip}\hline\noalign{\smallskip} \noalign{\smallskip}
         &    prot.           &                        &   \\
\noalign{\smallskip}
$3p_{1/2}$ & -0.375 &  0.153 & -0.222 \\
$3p_{3/2}$ & -0.371 &  0.152 & -0.219 \\
$2f_{5/2}$ & -0.278 &  0.168 & -0.110 \\
$1i_{13/2}$& -0.534 &  0.266 & -0.268 \\
$2f_{7/2}$ & -0.409 &  0.168 & -0.240 \\
$1h_{9/2}$ & -0.054 &  0.222 &  0.168 \\
$3s_{1/2}$ & -0.310 &  0.143 & -0.167 \\
$2d_{3/2}$ & -0.241 &  0.146 & -0.095 \\
$1h_{11/2}$& -0.017 &  0.246 &  0.229 \\
$2d_{5/2}$ &  0.435 &  0.147 &  0.582 \\
$1g_{7/2}$ & -0.271 &  0.197 & -0.074 \\
\noalign{\smallskip}\hline\noalign{\smallskip}
\end{tabular}
\end{center}
\label{tab:Pb-tad}
\end{table}

\begin{table}
\caption{PC corrections to single-particle energies (MeV) in $^{40}$Ca.}
\begin{center}
\begin{tabular}{ccccccc}
\noalign{\smallskip}\hline\noalign{\smallskip}

$\lambda$ & $Z_{\lambda}$ & $\eps^{(0)}_{\lambda}$ & \qquad $\delta
\eps_{\lambda}$ & &
$\eps_{\lambda}$ & $\eps^{\rm exp}_{\lambda}$ \cite{exp}\\
          &               &                        &    $3^-$ &    $1^-$         &
                 &                                 \\
\noalign{\smallskip}\hline\noalign{\smallskip} \noalign{\smallskip}
         &               &                        & neutr. &
                 &                            &       \\
\noalign{\smallskip}
$1f_{5/2}$ & 0.947 & -2.124 & 0.197 &  0.321 & -1.634 & -2.65 \\
$2p_{1/2}$ & 0.934 & -3.729 & -0.500 &  0.133 & -4.072& -4.42 \\
$2p_{3/2}$ & 0.916 & -5.609 & -0.450 &  0.130 & -5.902& -6.42\\
$1f_{7/2}$ & 0.947 & -9.593 &  0.591 &  0.173 & -8.870& -8.36\\
$1d_{3/2}$ & 0.965 & -14.257& 0.271 &  0.267 & -13.738& -15.64\\
$2s_{1/2}$ & 0.930 & -15.780& -0.439 &  0.184 & -16.017& -18.11\\
$1d_{5/2}$ & 0.969 & -19.985&  0.478 &  0.224 & -19.305&  -21.27\\
\noalign{\smallskip}\hline\noalign{\smallskip}
\noalign{\smallskip}
         &               &                        & prot. &
                 &                            &       \\
\noalign{\smallskip}
$1f_{5/2}$ & 0.963 &  4.359 & 0.230 &  0.300 & 4.869 & 4.60 \\
$2p_{1/2}$ & 0.950 &  2.456 & -0.432 & 0.062 & 2.104& 2.38 \\
$2p_{3/2}$ & 0.966 &  0.936 & 0.027 & 0.091 &  1.050& 0.63\\
$1f_{7/2}$ & 0.960 & -2.678 & 0.777 & -0.198 & -2.122& -1.09\\
$1d_{3/2}$ & 0.966 & -7.264& 0.289 & 0.262 & -6.733& -8.33\\
$2s_{1/2}$ & 0.931 & -8.663&-0.457 & 0.170 & -8.931& -10.85\\
$1d_{5/2}$ & 0.969 &-12.856& 0.466 & 0.216 & -12.196&  -13.73\\

\noalign{\smallskip}\hline\noalign{\smallskip}
\end{tabular}
\end{center}
\label{tab:Ca40}
\end{table}

\begin{table}
\caption{PC corrections to single-particle energies (MeV) in
$^{48}$Ca.}
\begin{center}
\begin{tabular}{cccccccc}
\noalign{\smallskip}\hline\noalign{\smallskip}

$\lambda$ & $Z_{\lambda}$ & $\eps^{(0)}_{\lambda}$ &  & $\delta
\eps_{\lambda}$& &
$\eps_{\lambda}$ & $\eps^{\rm exp}_{\lambda}$ \cite{exp}\\
          &               &                        &    $3^-$ & $2^+$ &  $1^-$          &
                 &                                 \\

\noalign{\smallskip}\hline\noalign{\smallskip} \noalign{\smallskip}
         &               &                        & neutr. &
                 &                            &       \\
\noalign{\smallskip}
$1g_{9/2}$ & 0.796 &  0.836 &  0.438 & -0.069 &  0.068 &  1.184 & 0.45 \\
$1f_{5/2}$ & 0.164 & -0.508 &   -    &   -    &    -   & -0.508 & -1.20 \\
$2p_{1/2}$ & 0.773 & -3.890 & -0.095 & -0.457 &  0.098 & -4.241 & -3.12 \\
$2p_{3/2}$ & 0.939 & -5.784 & -0.116 & -0.068 &  0.119 & -5.846 & -5.15 \\
$1f_{7/2}$ & 0.965 & -9.488 &  0.153 &  0.095 &  0.121 & -9.132 & -9.95 \\

\noalign{\smallskip}\hline\noalign{\smallskip} \noalign{\smallskip}
         &               &                        & prot. &
                 &                            &       \\
\noalign{\smallskip}
$1f_{5/2}$ & 0.873 & -4.048 &  0.076 & -0.330 &  0.249 & -4.052 & -4.55 \\
$2p_{1/2}$ & 0.648 & -3.549 & -0.114 & -1.399 &  0.157 & -4.427 & -5.05 \\
$2p_{3/2}$ & 0.604 & -4.731 & -0.089 &  0.390 &  0.126 & -4.473 & -6.55 \\
$1f_{7/2}$ & 0.899 & -9.909 &  0.144 & -0.305 &  0.176 & -9.896 & -9.63 \\
$1d_{3/2}$ & 0.917 &-16.172 &  0.099 &  0.369 &  0.190 &-15.568 & -16.17\\
$2s_{1/2}$ & 0.915 &-15.098 & -0.024 &  0.476 &  0.147 &-14.550 & -15.81\\
$1d_{5/2}$ & 0.116 &-19.913 &    -   &     -  &    -   &-19.913 & -21.58\\

\noalign{\smallskip}\hline\noalign{\smallskip}

\end{tabular}
\end{center}
\label{tab:Ca48}
\end{table}

\begin{table}
\caption{PC corrections to single-particle energies (MeV) in $^{56}$Ni.}
\begin{center}
\begin{tabular}{cccccccc}
\noalign{\smallskip}\hline\noalign{\smallskip}

$\lambda$ & $Z_{\lambda}$ & $\eps^{(0)}_{\lambda}$ &   &$\delta
\eps_{\lambda}$ & &
$\eps_{\lambda}$ & $\eps^{\rm exp}_{\lambda}$ \cite{exp} \\
          &               &                        &    $3^-$ & $2^+$ &  $1^-$          &
                 &                                 \\
\noalign{\smallskip}\hline\noalign{\smallskip} \noalign{\smallskip}
         &               &                        & neutr. &
                 &                            &       \\
\noalign{\smallskip}
$1g_{9/2}$ & 0.777 & -5.311 & -0.263 & -0.120 &  0.097 & -5.533 & -6.55 \\
$2p_{1/2}$ & 0.774 & -9.615 & -0.101 & -0.411 &  0.149 & -9.895 & -9.14 \\
$1f_{5/2}$ & 0.008 & -8.258 &  -     & -      &  -     & -8.258 & -9.48 \\
$2p_{3/2}$ & 0.933 &-11.064 & -0.111 & -0.042 &  0.126 &-11.089 & -10.25 \\
$1f_{7/2}$ & 0.945 &-15.588 &  0.309 &  0.130 &  0.137 &-15.044 & -16.65 \\
$1d_{3/2}$ & 0.927 &-20.763 &  0.424 &  0.141 &  0.148 &-20.103 & -19.84\\
$2s_{1/2}$ & 0.752 &-20.911 &  0.800 &  0.180 &  0.120 &-20.084 & -20.40\\
\noalign{\smallskip}\hline\noalign{\smallskip} \noalign{\smallskip}
         &               &                        & prot. &
                 &                            &       \\
\noalign{\smallskip}
$1g_{9/2}$ & 0.809 &  3.722 & -0.224 & -0.054 &  0.084 &  3.565 & 2.77 \\
$2p_{1/2}$ & 0.761 & -0.648 & -0.106 & -0.491 &  0.122 & -1.011 & 0.37 \\
$1f_{5/2}$ & 0.445 &  0.713 &  0.205 & -0.307 &  0.215 &  0.763 & 0.29 \\
$2p_{3/2}$ & 0.911 & -1.905 & -0.119 & -0.123 &  0.106 & -2.029 & -0.74 \\
$1f_{7/2}$ & 0.963 & -6.276 &  0.280 &  0.178 &  0.129 & -5.711 & -7.16 \\
$1d_{3/2}$ & 0.941 &-11.432 &  0.388 &  0.217 &  0.145 &-10.726 & -10.08\\
$2s_{1/2}$ & 0.815 &-11.349 &  0.659 &  0.101 &  0.111 &-10.639 & -10.72\\
\noalign{\smallskip}\hline\noalign{\smallskip}
\end{tabular}
\end{center}
\label{tab:Ni56}
\end{table}

\begin{table}
\caption{PC corrections to single-particle energies (MeV) in
$^{78}$Ni. * Experimental values are interpolated
from data for neighboring nuclei.}
\begin{center}
\begin{tabular}{cccccccc}
\noalign{\smallskip}\hline\noalign{\smallskip}
$\lambda$ & $Z_{\lambda}$ & $\eps^{(0)}_{\lambda}$ &   &$\delta
\eps_{\lambda}$ & &
$\eps_{\lambda}$ & $\eps^{\rm exp}_{\lambda}$ \cite{exp}* \\
          &               &                        &    $3^-$ & $2^+$ &  $1^-$          &
                 &                                 \\
\noalign{\smallskip}\hline\noalign{\smallskip} \noalign{\smallskip}
         &               &                        & neutr. &
                 &                            &       \\
\noalign{\smallskip}
$3s_{1/2}$ & 0.873 & -1.045 & -0.080 & -0.409 &  0.017 & -1.457 & -1.44 \\
$2d_{5/2}$ & 0.915 & -1.477 & -0.040 & -0.162 &  0.052 & -1.615 & -1.98 \\
$1g_{9/2}$ & 0.918 & -5.481 &  0.169 &  0.264 &  0.068 & -5.021 & -5.86 \\
$2p_{1/2}$ & 0.910 & -8.268 & -0.059 &  0.349 &  0.083 & -7.929 & -7.21 \\
$1f_{5/2}$ & 0.912 & -8.553 &  0.172 &  0.364 &  0.114 & -7.960 & -8.39 \\
$1p_{3/2}$ & 0.724 & -9.641 &  0.446 &  0.378 &  0.054 & -9.005 & -8.54 \\
\noalign{\smallskip}\hline\noalign{\smallskip} \noalign{\smallskip}
         &               &                        & prot. &
                 &                            &       \\
\noalign{\smallskip}
$1g_{9/2}$ & 0.773 &-11.138  &-0.190 & -0.152 & 0.099 &-11.326 & -8.91 \\
$2p_{1/2}$ & 0.679 &-14.185  &-0.104 & -0.811 & 0.125 &-14.721 & -12.04\\
$2p_{3/2}$ & 0.880 &-15.526  &-0.115 & -0.161 &  0.102 &-15.680 & -13.44 \\
$1f_{5/2}$ & 0.927 &-15.061  & 0.168 & -0.081 &  0.125 & -14.864 & -14.94 \\
$1f_{7/2}$ & 0.943 &-20.245  & 0.214 &  0.195 &  0.112 &-19.754 & -20.06 \\
\noalign{\smallskip}\hline\noalign{\smallskip}
\end{tabular}
\end{center}
\label{tab:Ni78}
\end{table}

\begin{table}
\caption{PC corrections to single-particle energies (MeV) in
$^{100}$Sn. * Experimental values  are
interpolated from data for neighboring nuclei.}
\begin{center}
\begin{tabular}{cccccccc}
\noalign{\smallskip}\hline\noalign{\smallskip}
$\lambda$ & $Z_{\lambda}$ & $\eps^{(0)}_{\lambda}$ &   &$\delta
\eps_{\lambda}$ & &
$\eps_{\lambda}$ & $\eps^{\rm exp}_{\lambda}$ \cite{exp}*\\
          &               &                        &    $3^-$ & $2^+$ &  $1^-$          &
                 &                                 \\
\noalign{\smallskip}\hline\noalign{\smallskip} \noalign{\smallskip}
         &               &                        & neut. &
                 &                            &       \\
\noalign{\smallskip}
$1h_{11/2}$& 0.755 & -7.630 & -0.314 & -0.142 &  0.061 & -7.928  & -7.78 \\
$2d_{3/2}$ & 0.810 & -9.087 & -0.097 & -0.568 &  0.083 & -9.559  & -9.48 \\
$3s_{1/2}$ & 0.661 & -9.158 & -0.194 & -0.977 &  0.060 & -9.893  & -9.58 \\
$2d_{5/2}$ & 0.899 &-11.180 & -0.121 & -0.152 &  0.058 & -11.374 & -11.05 \\
$1g_{7/2}$ & 0.928 & -9.705 &  0.193 & -0.129 &  0.100 & -9.552  & -11.13 \\
$1g_{9/2}$ & 0.938 &-16.449 &  0.268 &  0.199 &  0.077 & -15.939 & -17.48 \\
$2p_{1/2}$ & 0.941 &-18.432 & -0.068 &  0.232 &  0.074 & -18.209 & -17.94 \\
\noalign{\smallskip}\hline\noalign{\smallskip} \noalign{\smallskip}
         &               &                        & prot. &
                 &                            &       \\
\noalign{\smallskip}
$1g_{7/2}$ & 0.930 &  4.077 & 0.206  & -0.132 &  0.097 &  4.237 & 4.00 \\
$2d_{5/2}$ & 0.908 &  2.812 & -0.136 & -0.143 &  0.044 &  2.599 & 3.10 \\
$1g_{9/2}$ & 0.938 & -2.345 &  0.256 &  0.196 &  0.072 & -1.853 & -2.86 \\
$2p_{1/2}$ & 0.942 & -4.081 & -0.072 &  0.221 &  0.068 & -3.877 & -3.65 \\
$2p_{3/2}$ & 0.750 & -5.360 &  0.647 &  0.275 &  0.050 & -4.631 & -4.75 \\
$1f_{5/2}$ & 0.891 & -6.030 &  0.276 &  0.299 & 0.074  & -5.451 & -8.99 \\
\noalign{\smallskip}\hline\noalign{\smallskip}
\end{tabular}
\end{center}
\label{tab:Sn100}
\end{table}

\begin{table}
\caption{PC corrections to single-particle energies (MeV) in $^{132}$Sn.}
\begin{center}
\begin{tabular}{cccccccc}
\noalign{\smallskip}\hline\noalign{\smallskip}

$\lambda$ & $Z_{\lambda}$ & $\eps^{(0)}_{\lambda}$ &   &$\delta
\eps_{\lambda}$ & &
$\eps_{\lambda}$ & $\eps^{\rm exp}_{\lambda}$\cite{exp} \\
          &               &                        &    $3^-$ & $2^+$ &  $1^-$          &
                 &                                 \\
\noalign{\smallskip}\hline\noalign{\smallskip} \noalign{\smallskip}
         &               &                        & neut. &
                 &                            &       \\
\noalign{\smallskip}
$1i_{13/2}$& 0.734 &  0.745 & -0.368 & -0.085 &  0.032 &  0.436  &  0.25 \\
$2f_{5/2}$ & 0.927 & -0.255 & -0.076 & -0.224 &  0.025 & -0.510  & -0.44 \\
$3p_{1/2}$ & 0.942 & -0.629 & -0.117 & -0.187 & -0.001 & -0.916  & -0.79 \\
$1h_{9/2}$ & 0.942 &  0.192 &  0.119 & -0.112 &  0.080 &  0.274 & -0.88 \\
$3p_{3/2}$ & 0.919 & -1.095 & -0.100 & -0.234 &  0.011 & -1.392  & -1.59 \\
$2f_{7/2}$ & 0.938 & -2.319 & -0.084 & -0.084 &  0.029 & -2.449 & -2.45 \\
$2d_{3/2}$ & 0.945 & -8.044 & -0.080 &  0.177 &  0.051 & -7.904 & -7.39 \\
$1h_{11/2}$ & 0.948 &  -7.472 &  0.215 &  0.135&   0.047&  -7.096 & -7.46 \\
$3s_{1/2}$ & 0.939 & -8.159 & -0.120 &  0.201 &  0.029 & -8.056 &  -7.72 \\
$2d_{5/2}$ & 0.727 & -9.993 &  0.619 &  0.206 &  0.031 & -9.371 &  -9.04 \\
$1g_{7/2}$ & 0.942 & -9.620 &  0.173 &  0.193 &  0.059 & -9.220 &  -10.28 \\
\noalign{\smallskip}\hline\noalign{\smallskip} \noalign{\smallskip}
         &               &                        & prot. &
                 &                            &       \\
\noalign{\smallskip}
$1h_{11/2}$ & 0.832 &  -7.056 &  -0.174 &  -0.044 &   0.056 &  -7.190 & -6.86 \\
$2d_{3/2}$ & 0.858 & -7.606 &  -0.104 &  -0.304 &   0.065 &  -7.900 & -6.95 \\
$2d_{5/2}$ & 0.921 & -9.420 &  -0.153 &  -0.063 &   0.048 &  -9.576 & -8.69 \\
$1g_{7/2}$ & 0.967 & -9.892 &   0.182 &  -0.010 &   0.063 &  -9.665 & -9.65 \\
$1g_{9/2}$ & 0.963 &-14.842 &   0.221 &   0.094 &   0.062 & -14.479 & -15.78\\
$2p_{1/2}$ & 0.963 &-16.073 &  -0.059 &   0.100 &   0.052 & -15.983 & -16.13 \\
\noalign{\smallskip}\hline\noalign{\smallskip}
\end{tabular}
\end{center}
\label{tab:Sn132}
\end{table}

\begin{table*}
\caption{PC corrections to single-particle energies (MeV) in $^{208}$Pb.}
\begin{center}
\begin{tabular}{ccccccccccc}
\noalign{\smallskip}\hline\noalign{\smallskip}
$\lambda$ & $Z_{\lambda}$ & $\eps^{(0)}_{\lambda}$ &   & & $\delta
\eps_{\lambda}$& & &
$\eps_{\lambda}$ & $\eps^{\rm exp}_{\lambda}$ \cite{exp} \\
          &               &                        &    $3^-$ & $5^-_1$ &$2^+_1$ & $\sum_{\rm rest} $  &$1^-$          &
                 &                                 \\
\noalign{\smallskip}\hline\noalign{\smallskip} \noalign{\smallskip}
         &               &                        & neutr. &
                 &                            &       \\
\noalign{\smallskip}
$3d_{3/2}$ & 0.879 & -0.709 & -0.137 & -0.027 & -0.086 & -0.278 &  0.004 & -1.171 & -1.40 \\
$2g_{7/2}$ & 0.886 & -1.091 & -0.081 & -0.013 & -0.095 & -0.215 &  0.026 & -1.426 & -1.45 \\
$4s_{1/2}$ & 0.895 & -1.080 & -0.118 & -0.028 & -0.066 & -0.240 &  0.003 & -1.483 & -1.90\\
$3d_{5/2}$ & 0.873 & -1.599 & -0.124 & -0.034 & -0.104 & -0.234 &  0.009 & -2.023 & -2.37 \\
$1j_{15/2}$& 0.618 & -2.167 & -0.504 & -0.016 & -0.025 &  0.009 &  0.025 & -2.483 & -2.51 \\
$1i_{11/2}$& 0.945 & -2.511 &  0.140 &  0.022 & -0.030 &  0.023 &  0.041 & -2.327 & -3.16 \\
$2g_{9/2}$ & 0.882 & -3.674 & -0.167 & -0.005 & -0.032 & -0.097 &  0.018 & -3.924 & -3.94 \\
$3p_{1/2}$ & 0.926 & -7.506 & -0.167 & -0.033 &  0.074 &  0.058 &  0.022 & -7.549 & -7.37\\
$2f_{5/2}$ & 0.923 & -8.430 & -0.092 & -0.006 &  0.066 &  0.124 &  0.032 & -8.316 & -7.94 \\
$3p_{3/2}$ & 0.913 & -8.363 & -0.149 &  0.004 &  0.081 &  0.074 &  0.016 & -8.338 & -8.27 \\
$1i_{13/2}$& 0.902 & -9.411 &  0.269 &  0.052 &  0.054 &  0.154 &  0.032 & -8.905 & -9.00 \\
$2f_{7/2}$ & 0.567 &-10.708 &  0.815 &  0.023 &  0.098 &  0.190 &  0.020 &-10.059 & -9.71 \\
$1h_{9/2}$ & 0.892 &-11.009 &  0.184 &  0.021 &  0.070 &  0.223 &  0.033 &-10.535 & -10.78 \\
\noalign{\smallskip}\hline\noalign{\smallskip} \noalign{\smallskip}
         &               &                        & prot. &
                 &                            &       \\
\noalign{\smallskip}
$3p_{1/2}$ & 0.005 &  0.484  &   -     &  -      & -       &    -    &  -      &   0.484 & -0.17\\
$3p_{3/2}$ & 0.690 &  -0.249 &  -0.219 &  -0.100 &  -0.154 &  -0.365 &   0.026 &  -0.810 & -0.68 \\
$2f_{5/2}$ & 0.812 &  -0.964 &  -0.110 &  -0.016 &  -0.106 &  -0.248 &   0.036 &  -1.325 & -0.98\\
$1i_{13/2}$& 0.741 &  -2.082 &  -0.268 &   0.012 &  -0.021 &   0.039 &   0.034 &  -2.234 & -2.19 \\
$2f_{7/2}$ & 0.859 &  -3.007 &  -0.240 &  -0.014 &  -0.013 &  -0.095 &   0.025 &  -3.298 & -2.90 \\
$1h_{9/2}$ & 0.958 &  -4.232 &   0.168 &   0.023 &   0.007 &   0.052 &   0.035 &  -3.959 & -3.80 \\
$3s_{1/2}$ & 0.929 &  -7.611 &  -0.167 &   0.018 &   0.048 &   0.051 &   0.026 &  -7.633 & -8.01 \\
$2d_{3/2}$ & 0.937 &  -8.283 &  -0.095 &   0.006 &   0.052 &   0.068 &   0.031 &  -8.223 & -8.36\\
$1h_{11/2}$& 0.931 &  -8.810 &   0.229 &   0.021 &   0.020 &   0.134 &   0.037 &  -8.399 & -9.36 \\
$2d_{5/2}$ & 0.711 &  -9.782 &   0.582 &   0.006 &   0.043 &   0.113 &   0.024 &  -9.234 & -9.70 \\
$1g_{7/2}$ & 0.423 & -11.735 &  -0.074 &   0.056 &   0.087 &   0.190 &   0.029 & -11.613 & -11.49 \\
\noalign{\smallskip}\hline\noalign{\smallskip}
\end{tabular}
\end{center}
\label{tab:Pb208}
\end{table*}

\begin{table}
\caption{PC effect on average deviations  $\langle\delta \eps_{\lambda}\rangle_{\rm rms}$ (MeV) of
the theory predictions for single-particle energies from the
experimental values for the DF3-a functional.}
\begin{center}
\begin{tabular}{cccc}
\noalign{\smallskip}\hline\noalign{\smallskip}

Nucleus   & $N$ & DF3-a+ph  & DF3-a \\
\noalign{\smallskip}\hline\noalign{\smallskip}

$^{40}$Ca & 14 & 1.30 &  1.25  \\

$^{48}$Ca & 12 & 1.05 & 1.00  \\

$^{56}$Ni & 14 & 0.98 &0.97  \\

$^{78}$Ni & 11 & 1.34 & 1.41 \\

$^{100}$Sn& 13 & 1.21 & 1.17  \\

$^{132}$Sn& 17 & 0.63 & 0.66   \\

$^{208}$Pb& 24 & 0.38 & 0.51  \\

\noalign{\smallskip}\hline\noalign{\smallskip}

total & 105 & 0.97 & 0.98  \\

\noalign{\smallskip}\hline\noalign{\smallskip}
\end{tabular}
\end{center} \label{tab:rmsPC}
\end{table}

\begin{table}
\caption{Single-particle energies (MeV) with PC corrections in $^{208}$Pb.
Comparison with predictions of the RMF theory \cite{Litv-Ring}.}
\begin{center}
\begin{tabular}{cccc}
\noalign{\smallskip}\hline\noalign{\smallskip}
$\lambda$ &
$\eps_{\lambda}[{\rm DF3{-}a{+}ph}]$ & $\eps^{\rm exp}_{\lambda}$ \cite{exp}& $\eps_{\lambda}[{\rm RMF {+} ph}]$ \\
 \noalign{\smallskip}\hline\noalign{\smallskip} \noalign{\smallskip}
         &             neutr. &    &       \\
\noalign{\smallskip}
$3d_{3/2}$ &  -1.171 & -1.40 &  -0.63 \\
$2g_{7/2}$ &  -1.426 & -1.45 &  -1.14 \\
$4s_{1/2}$ &  -1.483 & -1.90 &  -0.92 \\
$3d_{5/2}$ &  -2.023 & -2.37 &  -1.39 \\
$1j_{15/2}$&  -2.483 & -2.51 &  -1.84 \\
$1i_{11/2}$&  -2.327 & -3.16 &  -3.30 \\
$2g_{9/2}$ &  -3.924 & -3.94 &  -3.29 \\

$3p_{1/2}$ & -7.549 & -7.37 & -7.68 \\
$2f_{5/2}$ & -8.316 & -7.94  & -8.66 \\
$3p_{3/2}$ & -8.338 & -8.27 & -8.26 \\
$1i_{13/2}$& -8.905 & -9.00 & -9.10 \\
$2f_{7/2}$ & -10.059 & -9.71 & -9.71 \\
$1h_{9/2}$ & -10.535 & -10.78& -11.96 \\
\noalign{\smallskip}\hline\noalign{\smallskip} \noalign{\smallskip}
         &               prot. &   &       \\
\noalign{\smallskip}
$3p_{1/2}$ &  0.484 & -0.17 &  1.09 \\
$3p_{3/2}$ & -0.810 & -0.68 & -0.16 \\
$2f_{5/2}$ & -1.325 & -0.98 & -1.07 \\
$1i_{13/2}$& -2.234 & -2.19 & -2.49 \\
$2f_{7/2}$ & -3.298 & -2.90 & -2.87 \\
$1h_{9/2}$ & -3.959 & -3.80 & -5.04 \\

$3s_{1/2}$ & -7.633 & -8.01 & -8.41 \\
$2d_{3/2}$ & -8.223 & -8.36 & -9.33 \\
$1h_{11/2}$& -8.399 & -9.36 & -9.92 \\
$2d_{5/2}$ & -9.234 & -9.70 & -10.05 \\
$1g_{7/2}$ & -11.613 & -11.49 & -13.74 \\
\noalign{\smallskip}\hline\noalign{\smallskip}
$\langle\delta \eps_{\lambda}\rangle_{\rm rms}$ & 0.38 & &0.81 \\
\noalign{\smallskip}\hline
\end{tabular}
\end{center}
\label{tab:Pb208-compar}
\end{table}

The $L$-phonon excitation energies $\omega_L$ and creation
amplitudes $g_L({\bf r})$ were found by solving the self-consistent
Eq. (\ref{g_L})  with the DF3-a functional. In more detail, the
procedure is described in \cite{BE2}. The results for $\omega_L$ and
$B(EL)$ values are listed in Table \ref{tab:BEL}. All the $L$-phonons
we consider are the surface vibrations which belong to  the
Goldstone mode corresponding to the spontaneous breaking of the
translation symmetry in nuclei \cite{KhS}. The coordinate form of
their creation amplitudes $g_L({\bf r})$ is very close to that for
the ghost phonon which is the lowest energy member of this mode:
\beq g_L(r)=\alpha_L \frac {dU} {dr} +\chi_L(r), \label{gLonr}\eeq
where the in-volume correction $\chi_L(r)$ is rather small. The first,
surface term on the right-hand sight. of Eq. (\ref{gLonr}) corresponds to the
BM model for the surface vibrations \cite{BM2}, the amplitude
$\alpha_L$ being related to the dimensionless BM amplitude $\beta_L$
as follows: $\alpha_L=R \beta_L$, where $R=r_0 A^{1/3}$ is the
nucleus radius, and $r_0=1.2\;$fm.

The smallness of the in-volume component $\chi_L$ is demonstrated in
Fig. \ref{fig:g3mi} for the $3^-_1$ state in $^{208}$Pb, which is the
most collective one among the surface vibrations in this nucleus.
The small spin components $S=1$ are also displayed.  To make them
distinguishable, they are multiplied by the factor of 10.  The
smallness of the spin components is typical for $L$-phonons with
a high collectivity. For phonons which are less collective, e.g.,
the $2^+_1$ state in $^{208}$Pb, the spin component is more important
and should be taken into account. In any case, we always took it
into account for all phonons.

Similarly,  the surface component also dominates in the transition
density: \beq \rho_L(r)=\alpha_L \frac {d\rho} {dr} +\eta_L(r).
\label{gLonr1}\eeq

If one neglects in-volume contributions, the tadpole PC term (\ref{tad}) can be reduced
to a form similar to (\ref{tad}):
\beq
\delta\Sigma^{\rm tad}_L = \frac {\alpha_L ^2} 2 \frac {2L+1} 3
\triangle U(r). \label{tad-L}\eeq As demonstrated in
\cite{Platon}, the in-volume corrections to Eq. (\ref{tad-L}) are,
indeed, small for heavy nuclei, e.g., for $^{208}$Pb. At the same
time, for light nuclei, e.g., $^{40,48}$Ca, the accurate solution
\cite{Platon} of Eq. (\ref{dgL})  diminishes the approximate value,
(\ref{tad-L}), for the tadpole term by $\simeq 30$\%.

Below we neglect the in-volume corrections for all nuclei
considered. To find the phonon amplitudes $\alpha_L$,
 we used the definition \beq \alpha_L^{\tau}= \frac {g_L^{\tau,{\rm max}}}
  {\left(\frac {dU} {dr}\right)^{\tau,{\rm max}} },     \label{alpL}\eeq with obvious notation.
It should be noted that the values of $\alpha_L^n$ and $\alpha_L^p$
are always very close to each other and to that which follows from the
BM model formula for $B(EL)$: $B(EL)_{\rm BM}= \left(
3Z/4\pi\right)^2\beta_L^2 R^{2L}$ \cite{BM2}, where the
dimensionless BM phonon creation amplitude $\beta_L$ related to that
used by us as $\alpha_L=\beta_L R /\sqrt{2L+1}$, $R=1.2\,A^{1/3}$.
For example, for the $3^-_1$ state in $^{208}$Pb we have:
$\alpha_L^n=0.32\;$fm, $\alpha_L^p=0.33\;$fm, and $\alpha_L^{\rm
BM}=0.30\;$fm.

Separate contributions of pole and tadpole terms for PC corrections
from the first $3^-$ state to single-particle levels for
$^{40}$Ca are listed in Table \ref{tab:Ca-tad},
 and those for  $^{208}$Pb in Table \ref{tab:Pb-tad}. The tadpole correction $\delta \eps^{\rm
tad}_{\lambda}$ is always positive, whereas the pole one $\delta
\eps^{\rm pole}_{\lambda}$ is, as a rule, negative. For such cases,
these two terms partially cancel each other. In the $^{40}$Ca nucleus,
these contributions are of the same order, and the sum  proves to be
positive in almost half of the cases. As mentioned
above, the tadpole values in Table \ref{tab:Ca-tad} could be reduced
by $\simeq 30$\%, providing the accurate solution \cite{Platon} of
Eq. (\ref{dgL}). In $^{208}$Pb, the role of the tadpole term is, on
average, smaller, but still important. In this case, the in-volume
corrections to Eq. (\ref{tad-L}) are small. Indeed, ``the surface-to-volume ratio'' 
decreases as $\propto A^{-1/3}$ for heavy nuclei,
therefore the surface vibrations resemble the modes of a classical
liquid drop, not penetrating inside its volume.

Consider now  the final results of the DF3-a functional for the
single-particle spectra for magic nuclei with inclusion of the PC
corrections. Let us begin with $^{40}$Ca, in Table \ref{tab:Ca40}. It
also contains the $Z_{\lambda}$-factors, which are used in the final
expression (\ref{eps-PC}) for the single-particle energy. The
difference  $1-Z_{\lambda}$ determines the scale of the PC effects.
The inequality $1-Z_{\lambda}\ll 1$ justifies the validity of the
perturbation theory in $g_L^2$. In addition to the $3^-$ state, 
Table \ref{tab:Ca40} lists the corrections due to the recoil effect from the
spurious $1^-$ state. For this nucleus, the latter is significant:
for several states, comparable with that from  the $3^-$ state. The agreement of the
PC corrections with the data is a little worse. Now  the
total average error is 1.30 MeV, compared to the 1.25 MeV without PC
corrections. The  main reason for this disagreement is the
overestimate of the tadpole term discussed above.

In $^{48}$Ca (Table \ref{tab:Ca48}) there are two states,
$1f_{5/2}^n$ and $1d_{5/2}^p$, with anomalously small values of
$Z_{\lambda}$. This occurs because of the occasional smallness of one
of the denominators in Eq. (\ref{dSig2}) due to some semi-generation
of the energies $\eps_{\lambda}$ and $\eps_{\lambda}\pm \omega_L$.
Of course, in this situation the plain perturbation theory is not
valid. An improved approach should be developed with exact
diagonalization of the ``two-level'' problem. Fortunately, such
cases are very rare: two  for $^{48}$Ca, one for $^{56}$Ni and one
for $^{208}$Pb. Therefore  we postpone the solution of this problem
skipping the calculation of the energy corrections $\delta
\eps_{\lambda}$ for these states. They are reported in Tables
\ref{tab:Ca48}, \ref{tab:Ni56} and \ref{tab:Pb208} to call attention
to this problem. For  the $^{48}$Ca  nucleus the phonon $2^+$ is added,
as sometimes  its contribution exceeds that of the $3^-$ state. The
contribution of the recoil effect is less than in $^{40}$Ca but
still important. It should be noted that a rough estimate of this
effect is $\simeq \eps_{\rm F}/A$, so it becomes small for heavy
nuclei, but, as a rule, not negligible. We have included it for all
nuclei, as it is, in fact, model independent.

For all nuclei from $^{56}$Ni to $^{132}$Sn (Tables \ref{tab:Ni56}
-- \ref{tab:Sn132}), the set of phonons we take into account is the
same as for $^{48}$Ca, i.e. $3^-,2^+$ and the ghost $1^-$ state. For
all of them the contributions of the $3^-$ and $2^+$ phonons are of the same
order of magnitude, whereas the $1^-$ contribution diminishes in
accordance with the above estimate. For  $^{208}$Pb we calculated
the contributions of nine phonons, $3^-$, $5^-_{1,2}$, $2^+_{1,2}$,
$4^+_{1,2}$, and $2^+_{1,2}$. As a rule, the contribution of the
$3^-$-phonon dominates. However, sometimes the contribution of all
other phonons is comparable with that of $3^-$. For this nucleus,
the PC corrections improve the description of the single-particle
spectrum. The average error is now 0.38 MeV instead of 0.51 MeV.

The average deviations $<\delta \eps_{\lambda}>_{\rm rms}$ for all
nuclei we consider for the DF3-a functional, with and without PC
corrections, are presented in Table \ref{tab:rmsPC}. For lighter
nuclei, $A=40\div 100$, PC corrections worsen the agreement a bit,
with the only exception of $^{78}$Ni.  For heavy nuclei, $^{132}$Sn
and $^{208}$Pb, the agreement becomes better.

To conclude this section, we compare in Table
\ref{tab:Pb208-compar} our results for $^{208}$Pb with predictions
of the RMF theory \cite{Litv-Ring}, the only calculation we know
where PC corrections to the single-particle spectrum are found
self-consistently. In this calculation only the pole diagram in Fig.
15 is taken into account. It is seen that the agreement of our result
with the data is significantly better. For the RMF spectrum the
average deviation from the data is $\langle\delta
\eps_{\lambda}\rangle_{\rm rms}=0.81\,$MeV, which is two times worse
than the result of the DF3-a functional with PC corrections.

\section{Conclusion}

Single-particle spectra of seven magic nuclei, from $^{40}$Ca to
$^{208}$Pb, some of which have become available recently \cite{exp},
are described within the EDF method of Fayans {\it et al.}
Comparison is made with the predictions of the SHF method with the
functional HFB-17, the the record holder in describing nuclear
masses  among self-consistent approaches. Three versions of the
Fayans  functional are used, DF3 \cite{Fay3} and two options DF3-a,b, 
with different spin-orbit parameter values. One of these, DF3-a, was
suggested in \cite{Tol-Sap} to describe nuclei heavier than
lead. The second option, DF3-b,
 is found in this paper to give a better description of the spin-orbit
differences $\Delta_{nls}$. The bulk of the data \cite{exp} provides 35
such differences, which makes it possible to find the optimal set of
spin-orbit parameters. The DF3-b set is the most successful: the
average deviation from experimental values $\langle\delta
\Delta_{nls}\rangle_{\rm rms}$  is equal to 0.60 MeV. For
comparison, it is 0.68 MeV for  the DF3-a functional, about 1 MeV for the
original DF3 functional, and  more than 2 MeV for the HFB-17
functional.

Description of the single-particle energies for all three versions
 of the DF3 functional is also significantly better than for the HFB-17 functional.
To compare the accuracy of different theories, on average, we found
the average differences  $\langle\delta \eps_{\lambda}\rangle_{\rm rms}$
between theoretical and experimental values of the single-particle
energies $\eps_{\lambda}$. These quantities are found for each
nucleus and for the whole set of 105 levels. For each of the nuclei
under consideration the predictions of the Fayans functional are more
accurate. For example, for $^{40}$Ca, $\langle\delta
\eps_{\lambda}\rangle_{\rm rms}$ values are equal to (1.08 -- 1.35)
MeV for the three versions of the DF3 functional and 1.64 MeV for the
HFB-17 functional. For the $^{208}$Pb nucleus, the advantage of the
Fayans functional is even more pronounced; the corresponding values
of $\langle\delta \eps_{\lambda}\rangle_{\rm rms}$ are (0.43 --
0.51) MeV for the DF3 functionals and 1.15 MeV for the HFB-17 one.  As
for the overall values of $\langle\delta \eps_{\lambda}\rangle_{\rm
rms}$, they are equal to 0.89 MeV for the DF3 functional, 0.98 MeV for
the DF3-a functional, and 0.89 MeV for the DF3-b. For the HFB-17 functional it is
equal to 1.40 MeV.

Thus, all three versions of the DF3 functional  describe the
single-particle levels with an accuracy of, on average, better than 1 MeV.
 We explain this by two main features of the Fayans EDF. First, the Fayans EDF
uses the bare mass, $m^*=m$, prescription of the Kohn--Sham method.
The self-consistent TFFS \cite{KhS} -- which takes into account not
only the momentum dependence, as does the SHF method, but also the energy dependence
effects -- leads  to a result which is rather close to the
Kohn--Sham prescription. This occurs due to the strong, almost-exact
cancellation  of the so-called $k$-mass and $E$-mass. The latter appears
due to the energy dependence of the quasiparticle mass operator on
energy, which has no analog in the SHF method. Second, the density
dependence of the Fayans EDF is much more sophisticated than that of the
SHF one. This is also an implicit consequence of the energy dependence
effects taken into account in the TFFS. In our opinion, the reason why
the HFB-17 functional, which describes nuclear masses perfectly
well, is less accurate for single-particle spectra
 is that the density dependence of SHF functionals is oversimplified for
describing more delicate nuclear characteristics.

The self-consistent description of the PC corrections to
single-particle spectra in magic nuclei is another subject of this
paper. Calculations are carried out for the DF3-a functional, which was
successful in describing the excitation energies and $B(E2)$ values
\cite{BE2,BE2-Web}, as well as quadrupole moments 
\cite{QEPJ,QEPJ-Web} in semimagic nuclei. The method developed in
\cite{KhS,Platon} is used, which permits us to calculate PC
contributions not only from the usual pole diagram but also from the
tadpole one. The latter is taken into account approximately, with
the anzatz (\ref{tad-L}), which neglects the in-volume components of
the vertices $g_L(r)$  of the surface vibrations. As shown in
\cite{Platon}, this approximation works well for heavy nuclei but it
is questionable for lighter ones. The tadpole contribution is
almost always positive as long as the pole contribution is usually negative.
As a result, the two terms, pole and tadpole, usually cancel each
other and the absolute value of the sum  is less than that from the
pole diagram alone. The contribution to $\eps_{\lambda}$ from the
spurious $1^-$ state, which describes the recoil effect due to the CM
motion, is also taken into account. It is very important for lighter
nuclei but rather minor for $^{208}$Pb.   After accounting for the PC
effects the average description of single-particle spectra becomes a
little worse for light nuclei but definitely  better for heavy
nuclei. For example, for $^{208}$Pb we obtained an average error equal to
0.38 MeV, versus of 0.51 MeV without PC corrections. As for overall
accuracy, the deviations of the theoretical predictions for
single-particle energies $\eps_{\lambda}$ from the experimental values
$\langle\delta \eps_{\lambda}\rangle_{\rm rms}$ averaged over more
than 100 states are 0.97 MeV and 0.98 MeV with and without
PC corrections, respectively. To improve the accuracy for light
nuclei, it is necessary to find the tadpole termtaking
into account exactly the in-volume contributions.

\section{Acknowledgment}

We are thankful to Jacek Dobaczewski for useful comments.
 
 The work was partly supported  by Grant No.
NSh-932.2014.2 from the Russian Ministry for Science and Education,
and by RFBR Grants Nos. 12-02-00955-a, 13-02-00085-a,
13-02-12106-ofi\_m, 14-02-00107-a, and 14-02-31353-mol\_a and Grant
IN2P3-RFBR under Agreement No. 110291054.

{}

\end{document}